\documentclass[journal]{IEEEtran}
\IEEEoverridecommandlockouts
\usepackage{cite}
\usepackage{amsmath,amsthm,amssymb,amsfonts}
\usepackage{array}
\usepackage{graphicx}
\usepackage{textcomp}
\usepackage{bbold}
\usepackage[mathscr]{eucal}
\usepackage[table]{xcolor}
\usepackage{xcolor}
\usepackage{cleveref}
\crefformat{section}{\S#2#1#3}
\usepackage[framemethod=tikz]{mdframed}
\usepackage[ruled,vlined,linesnumbered]{algorithm2e}
\usepackage{algpseudocode}
\usepackage{multirow}
\usepackage{rotating}
\newtheorem{assumption}{Assumption}

\newtheorem{example}{Example}
\usepackage[font=footnotesize]{caption}
\usepackage{subcaption}
\usepackage{pifont}
\usepackage[flushmargin]{footmisc}

\algnewcommand\LeftComment[2]{%
\hspace{#1\algindent}$\triangleright$ \eqparbox{COMMENT}{#2} \hfill %
}

\let\oldnl\nl% Store \nl in \oldnl
\newcommand{\nonl}{\renewcommand{\nl}{\let\nl\oldnl}}

\def\BibTeX{{\rm B\kern-.05em{\sc i\kern-.025em b}\kern-.08em
    T\kern-.1667em\lower.7ex\hbox{E}\kern-.125emX}}
\begin{document}
\title{Resource Aware Clustering for Tackling the Heterogeneity of Participants in Federated Learning}
\author{\IEEEauthorblockN{Rahul Mishra, Member IEEE, Hari Prabhat Gupta, Senior Member IEEE, and Garvit Banga}
\thanks{Rahul Mishra is with the Department of Information and Communication Technology, Dhirubhai Ambani Institute of Information and Communication Technology (DA-IICT), Gujrat, Gandhinagar, \\ e-mail: \textit{rahul\_mishra@daiict.ac.in}\\
Hari Prabhat Gupta is with the Department of Computer Science and Engineering, Institute of Technology (BHU) Varanasi, India, \\
e-mail: \textit{hariprabhat.cse@iitbhu.ac.in}\\
Garvit Banga is with the the Department of Metallurgical Engineering, Indian Institute of Technology (BHU) Varanasi, India, \\ e-mail: \textit{garvit.banga.met17@iitbhu.ac.in}}}

\maketitle
\begin{abstract}
Federated Learning is a training framework that enables multiple participants to collaboratively train a shared model while preserving data privacy and minimizing communication overhead. The heterogeneity of devices and networking resources of the participants delay the training and aggregation in federated learning. This paper proposes a federated learning approach to manoeuvre the heterogeneity among the participants using resource aware clustering. The approach begins with the server gathering information about the devices and networking resources of participants, after which resource aware clustering is performed to determine the optimal number of clusters using Dunn Indices. The mechanism of participant assignment is then introduced, and the expression of communication rounds required for model convergence in each cluster is mathematically derived. Furthermore, a master-slave technique is introduced to improve the performance of the lightweight models in the clusters using knowledge distillation. Finally, experimental evaluations are conducted to verify the feasibility and effectiveness of the approach and to compare it with state-of-the-art techniques.

\end{abstract}

\section{Introduction}
Federated Learning (FL) is a newly emerging paradigm that enables a distributed training framework where data collection and model training occur locally for each participant. Thus, it preserves data privacy and reduces communication overhead of transmitting data to the server~\cite{9917556}. Unlike traditional distributed training frameworks that require consensus after each local iteration, either through server or peer communication, FL minimizes the frequency of consensus among distributed participants. FL is initiated by the central server, which broadcasts a randomly initialized model to all participants. Each participant trains the received model using their local dataset and sends the Weight Parameter Matrices (WPM) to the server. The server then aggregates the WPM received from multiple participants and sends back the aggregated one, generating a robust and generalized model for each participant~\cite{Qu_2022_CVPR}.

FL participants exhibit significant heterogeneity in terms of devices and networking resources, including processing speed, available memory, and data transmission rate. Each participant uses its resources to load the model and train it locally. The availability of device resources among participants depends on their respective configurations and installed services, leading to irregular intervals between WPM generation. Furthermore, the data transmission rate affects the time required to upload WPM from participants to the server. Consequently, participant heterogeneity hinders the simultaneous transmission and aggregation of WPM. In other words, slower participants (i.e., stragglers) delay the entire training process. The server can mitigate this issue by setting a Maximum Allowable Response (MAR) time for training to minimize the delay caused by stragglers. However, using a fixed MAR time can result in inadequate training due to a reduced number of local updates across communication rounds on stragglers.

Previous researches on FL have addressed the issue of heterogeneity among participants by excluding stragglers from the training process~\cite{10.1145/3369583.3392686, clusterfl, xie2021multi}. However, removing stragglers from the training process deprives the system of their valuable datasets, which in turn reduces the model's generalization ability. Additionally, it also prevents the potential performance improvement of these stragglers using FL. Cluster-based techniques have been proposed in prior studies to address the heterogeneity among participants in FL. These techniques utilize the relationship between local datasets~\cite{clusterfl,xie2021multi}, the similarity of local updates~\cite{9207469}, and social relationships between participants~\cite{9459932} to form clusters. Nevertheless, these studies did not consider the devices and networking resources of participants during clustering. In their work~\cite{horvath2021fjord}, the authors pointed out the issue of heterogeneous devices in FL that restricts the size of the global model to accommodate stragglers. Similarly, in~\cite{diao2020heterofl}, the authors proposed a technique called HeteroFL to handle variations in computational and communication resources by generating multiple sized models and selecting the best one for each participant. Despite these benefits, neither~\cite{horvath2021fjord} and \cite{diao2020heterofl} have addressed the issue of improving performance of the lightweight models used by participants with limited resources. Additionally, earlier works~\cite{hinton2015distilling, li2020few, 9151346} employed Knowledge Distillation (KD) to enhance the performance of the lightweight model by leveraging insights from the large-sized; but, these methods were confined to centralized training.

This paper presents a novel approach called Fed-RAC (short for \underline{Fed}erated learning with \underline{R}esource \underline{A}ware \underline{C}lustering) to address the negative impact of participant heterogeneity in Federated Learning. We investigate the effect of participant heterogeneity and determine an expression for the required communication rounds per cluster. Fed-RAC is also designed to estimate the error caused by inconsistent objective functions in the presence of heterogeneous devices and networking resources. In particular, we focus on investigating the following problem: \textit{"How can we achieve satisfactory performance while training local models on heterogeneous participants in FL within the given MAR?"} To this end, the major contributions and novelty of this work are as follows:\\
\noindent $\bullet$ \textit{Resource aware clustering:} The first contribution is to conduct resource-aware clustering for identifying the most suitable number of clusters based on the devices and networking resources available to the participants. The server first gathers information regarding the processing speed, data transmission rate, and available memory of all participants to create resource vectors. These vectors are then subjected to unit-based normalization to bring their values within the range of $[0,1]$. To determine the optimal number of clusters, the server calculates the Dunn Indices~\cite{6170593} among the normalized resource vectors of all participants. \\
\noindent $\bullet$ \textit{Participants assignment to the clusters:} The next contribution is the allocation of participants to the identified clusters, ensuring that the model training within each cluster is performed within a specified maximum allowable response time and communication rounds. Additionally, a mathematical analysis is carried out to derive the expression for the communication round and error caused by an inconsistent objective function in the presence of heterogeneous participants.\\
\noindent $\bullet$ \textit{Master-slave technique}: Further, our approach introduces the master-slave technique to enhance the performance of the generic model in low-configuration clusters (slaves) by leveraging the model of the highest configuration cluster (master). In this technique, the master model is initially trained, and then it guides the training of slave models using knowledge distillation to improve their performance. \\
\noindent $\bullet$ \textit{Experimental validation:} In the end, we conduct experimental evaluations to confirm the effectiveness of the Fed-RAC approach. We validate our proposed method by comparing it with existing baseline techniques~\cite{diao2020heterofl, li2020federated, mcmahan2017communication, lai2021oort}, using various evaluation metrics and established datasets~\cite{lecun1998gradient,anguita2013public, krizhevsky2009learning, shl2}. The results demonstrate that the proposed approach achieves better performance in the presence of heterogeneous participants.

\textbf{Paper Organization:} Section~\ref{related-work} provides an overview of the related literature. Section~\ref{problem} outlines the preliminary information and problem statement of our proposed approach. Section~\ref{approach} details the Fed-RAC approach. Section~\ref{evaluation} evaluates the performance of our approach, while Section~\ref{discuss} presents the discussion and future directions for this work. Finally, Section~\ref{conc} concludes the paper.

\section{Background and Motivation}\label{related-work}
In this section, we provide a description of prior studies that focus on the heterogeneity of participants, clustering in FL, and knowledge distillation to enhance performance. 

\noindent $\bullet$ \textit{Heterogeneous participants in FL:} FL involves a significant number of participant devices with varying resources, leading to degraded performance and increased convergence time when running the same model on all participants~\cite{10.1145/3369583.3392686, 9139873, 9355774, wang2019adaptive, 9488756}. In~\cite{10.1145/3369583.3392686}, the authors proposed a system that selects participants for global aggregation and simultaneously generates WPM. The system discards straggling participants from the aggregation. To account for the slower computational speed of stragglers, the authors in~\cite{9139873} proposed reducing the CPU frequency of faster participants in the federation. The authors in~\cite{horvath2021fjord} identified the problem of heterogeneous devices in FL, which limits the size of the global model to accommodate low-resource or slow participants. They proposed a dynamically adaptive approach to model size called ordered dropout, FjORD. In~\cite{lai2021oort}, the authors presented a specialized technique called Oort, which prioritizes participant selection in FL. The authors in~\cite{li2020federated} introduced a framework called FedProx to handle the issue of data/task heterogeneity in FL. FedProx used a proximal term to minimize the impact of local updates.

In previous studies, various mechanisms have been proposed to address the issue of stragglers in FL, including asynchronous~\cite{xie2019asynchronous, chen2019communication} and semi-synchronous~\cite{ma2021fedsa} global update approaches. The authors in~\cite{xie2019asynchronous} introduced an asynchronous algorithm to optimize the FL-based training for stragglers. The algorithm solved the local regularization to ensure convergence in finite time and performed a weighted average to update the global model. Similarly, the authors in~\cite{chen2019communication} introduced the mechanism of asynchronous learning and weighted temporal aggregation on participants and server, respectively. To overcome the problem of higher waiting time in the asynchronous global updates, the authors in~\cite{ma2021fedsa} introduced the semi-asynchronous mechanism, where the server aggregates the weight parameters from a set of participants as per their arrival order in each communication round. 

\noindent $\bullet$ \textit{Clustering in FL:} The prior studies utilized the relationship between local datasets~\cite{clusterfl,xie2021multi}, the similarity of local updates~\cite{9207469}, and social relationship between the participants~\cite{9459932} to form clusters in FL. Authors in~\cite{clusterfl} exploited the intrinsic relationship between local datasets of multiple participants and proposed a similarity-aware system, namely ClusterFL. The system generated various clusters based on the similarity among local datasets. Similar to~\cite{clusterfl}, the authors in~\cite{xie2021multi} created various groups of participants as per the similarity among their local datasets. The group formation led to a minor loss over all the participants and provided communication efficiency. In~\cite{9207469} authors introduced a modified FL approach, where hierarchical clustering is performed as per the similarity of local updates. The authors in~\cite{diao2020heterofl} introduced the technique to handle variation in computational and communication resources. They named the technique as HeteroFL. 

\noindent $\bullet$ \textit{KD based performance improvement:} The existing literature introduced various techniques to improve the performance of the lightweight model using a large-size model via KD~\cite{hinton2015distilling, mishra2017apprentice, zhou2017rocket, 9151346}. The concept of KD was first introduced by the authors in~\cite{hinton2015distilling}, where the knowledge of a large-size model (teacher) is utilized to improve the performance of the lightweight model (student). The authors in~\cite{mishra2017apprentice} proposed the concept of simultaneous training of scratch teacher and student, which provided a soft target of logits to estimate the distillation loss. Finally, the authors in~\cite{9151346} introduced the concept of pre-trained teacher and scratch teacher-guided KD technique to improve the performance of student.    

\noindent \textbf{Motivation:} We observed the following limitations in existing literature. Prior studies discarded stragglers from the training to cope up with the heterogeneity of the resources among participants in FL~\cite{10.1145/3369583.3392686, clusterfl, xie2021multi}. When the stragglers are discarded, their available local datasets are not utilized during training, which reduces the generalization ability of all the participants. \textcolor{black}{In addition, discarding slow participants hampered their performance improvement via FL. Reducing the processing power of the participant device during training of the model slowdown the aggregation process~\cite{9139873}. The asynchronous federated learning mechanisms~\cite{xie2019asynchronous, chen2019communication} demand the server to wait for stragglers, leading to significant waiting time. The semi-asynchronous global aggregation mechnism~\cite{ma2021fedsa} is more effective than synchronous, but it discards some participants in each communication round. Suppressing the communication round for aggregation~\cite{wang2019adaptive} also increases the stale models at participants. The existing work exploited clustering in FL but did not consider the devices and networking resources during clustering~\cite{9207469, 9459932, xie2021multi,clusterfl}. The prior studies~\cite{hinton2015distilling, mishra2017apprentice, zhou2017rocket, 9151346} helped in improving the performance of the lightweight model using knowledge from the large-size model. However, these techniques were limited to centralized training.}  

\noindent \textcolor{black}{\textit{In summary, the existing FL approaches avoid straggler devices during aggregation at the central server. The asynchronous global update leads to higher weighting time at the server. The existing clustering mechanism in FL did not consider the resources of the participants, like memory, processing speed, and communication channel. Additionally, the existing work on KD to improve the performance of lightweight models are limited to centralized training.}}

\section{Preliminaries and problem statement}\label{problem}
This section describes the terminologies and notations, followed by the problem of the heterogeneous participants. 

\subsection{Preliminaries}
This work considers  a set $\mathcal{P}$ of $N$ participants and a central server, where $\mathcal{P}=\{p_1,\cdots,p_N\}$. We consider a multi-class classification problem with a set $Q$ of $c$ classes, \textit{i.e.}, $Q=\{1,\cdots, c\}$. Each participant $p_i$ has local dataset $\mathcal{D}_i$ with $n_i$ number of instances and set of $Q$ classes, where $1\le i \le N$. Let $(\mathbf{x}_{ij}, y_{ij})$ denotes an instance of dataset $\mathcal{D}_i$, where $1\le j \le n_i$. During training the model on the participant $p_i$ learn the mapping between $\mathbf{x}_{ij}$ and $y_{ij}$, $\forall j\in \{1\le j \le n_i\}$, to build a classifier $\Pi_i$. The classifier recognizes the class label of unidentified instances in testing. Let $B_i$ denotes the batch-size used for training model on $p_i$. Further, let $\tau_i$ represents the number of Stochastic Gradient Descent (SGD) operations performed in one round of training on $p_i$. $\tau_i$ is estimated as: $\tau_i=\lfloor E n_i /B_i\rfloor$, where $E$ is the number of local epochs to train on $p_i$. We can change $B_i$ and $n_i$ to change $\tau_i$.

\subsection{FL with heterogeneous participants} 
FL begins with the generation and random initialization of a model at the central server that further broadcasts the initialized model to all the participants. Each participant $p_i$ receives and trains the model using local dataset $\mathcal{D}_i$ with $n_i$ instances, where $1\le i \le N$. $p_i$ performs training for $E$ number of local epochs on a batch size of $B_i$ over $n_i$ instances using SGD operations $\tau_i$. The participant minimizes the local loss function $\mathcal{L}_i(\mathbf{w}_i)$, where $\mathbf{w}_i$ is the WPM of $p_i$. $\mathcal{L}_i(\mathbf{w}_i)$ is estimated as: $\mathcal{L}_i(\mathbf{w}_i)=\frac{1}{n_i}\sum_{j\leftarrow 1}^{n_i} \mathcal{L}_{ij}(\mathbf{w}_{ij})$, where  $\mathbf{w}_{ij}\in \mathbf{w}_i$, $1\le j \le n_i$, and $1\le i \le N$. The participant transfers estimated $\mathcal{L}_i(\cdot)$ and $\mathbf{w}_i$ to the server for global aggregation. Upon receiving local loss and WPM from all the participants, the server estimates global loss ($\mathcal{L}(\mathbf{w})$) and WPM ($\mathbf{w}$) as: 
\begin{align}\small\nonumber
\hspace{-0.3cm}\mathcal{L}(\mathbf{w})=\sum_{i\leftarrow 1}^{N} \Big(\frac{n_i}{n_1\cdots n_N}\Big)\mathcal{L}_i(\mathbf{w}_i), \mathbf{w}=\sum_{i\leftarrow 1}^{N}& \Big(\frac{n_i}{n_1\cdots n_N}\Big)\mathbf{w}_i.
\end{align}

The server broadcasts $\mathbf{w}$ for next round of training. The process of local training and aggregation are orchestrated for $\mathcal{R}$ iterations to achieve a trained model for all the participants. At each global iteration $t\in \mathcal{R}$ the local loss function and WPM are denoted as $\mathcal{L}_i^t(\mathbf{w}_i^t)$ and $\mathbf{w}_i^t$ for participant $p_i$, respectively, where $1\le i \le N$. $\mathbf{w}_i^t$ at global iteration $t$ ($t\in \mathcal{R}$) of participant $p_i$ is updated as:  $\mathbf{w}_{i}^t = \mathbf{w}_{i}^{t-1}-\eta \nabla\mathcal{L}_i^t(\mathbf{w}_i^t)$, where $\eta$ is the learning rate. Using above equation, we can define the objective function of FL as follows:
\begin{equation}\small\label{e3}
 \min_{\mathbf{w}^{\mathcal{R}}}\mathcal{L}(\mathbf{w}^{\mathcal{R}})=\sum_{t\leftarrow 1}^{\mathcal{R}}\sum_{i\leftarrow 1}^{N} \Big(\frac{n_i}{n_1+\cdots+n_N}\Big)\mathcal{L}_i^t(\mathbf{w}_i^t).
\end{equation}

\subsubsection{Heterogeneous participants} The heterogeneous participants in FL require non-identical training and communication time. Let $T_i$ denotes training and communication time of $p_i$, estimated as:  $T_i = T_i^a . E + T_i^c, \forall i \in \{1,2,\cdots N\}$, where $T_i^a$ is the training time for one local epoch, and $T_i^c$ is the per-round communication time for sharing WPM from $p_i$ to server. The participants train local model and communicate WPM in parallel. Thus, for each iteration $t\in \mathcal{R}$ the training and communication time $T^t$ depends on the slowest participant, where $T^t=\max_{1\le i \le N} \{T_i\}$. We obtain total training time, denoted as $\mathbb{T}(N,E,\mathcal{R})$, as: 
\begin{equation}\label{e5}
 \mathbb{T}(N,E,\mathcal{R}) = \sum_{t\leftarrow 1}^\mathcal{R} T^t = \sum_{t\leftarrow 1}^\mathcal{R} \max_{1\le i \le N} \{T_i\}.
\end{equation}

\subsubsection{Objective inconsistency}\label{obj} 
The server has a fixed MAR time to complete the global iterations, which reduces the training delay due to slow processing and communication of stragglers. 
It also minimizes the idle time of faster participants. However, the number of local SGD operations vary over heterogeneous participants within the fixed MAR time. The faster participants perform more local updates in contrast with stragglers. In addition, the number of local updates on the participants also vary across the communication rounds. The objective function of FL given in \eqref{e3} relies upon the assumptions that the number of local updates, $\tau_i$ for $p_i$ $\forall i\in\{1,2,\cdots, N\}$, remain same for all participants ($\tau_i=\tau$). However, the variation in the local updates on the heterogeneous participants results in an inconsistent objective function for FL~\cite{objective}. Let $\bar{\mathcal{L}}(\bar{\mathbf{w}}^{\mathcal{R}})$ denotes the inconsistent objective function, where $\bar{\mathbf{w}}^{\mathcal{R}}$ is the aggregated WPM generated after $\mathcal{R}$ global iterations. The error ($err$) between actual and inconsistent objective function is defined as: $err=|\bar{\mathcal{L}}(\bar{\mathbf{w}}^{\mathcal{R}})-\mathcal{L}(\mathbf{w}^{\mathcal{R}})|$. Thus, it is required to minimize error to mitigate the negative impact of heterogeneous participants.   

\begin{example}
Let $p_1-p_{10}$ denote $10$ participants in FL. In the absence of information about available resources, we can assume homogeneous participants with an equal number of data instances to estimate their local loss functions. It implies $\mathcal{L}_1(\mathbf{w}_1)=\mathcal{L}_2(\mathbf{w}_2)=\cdots=\mathcal{L}_{10}(\mathbf{w}_{10})=l_1$. Let $l_1=0.027$; thus, the objective function given in \eqref{e3} attains the value of $\mathcal{L}(\mathbf{w}^{\mathcal{R}})=0.027$. However, the participants may be heterogeneous; thus, the inconsistent objective function may obtain  $0.036$, which implies $err=0.009$. 
\end{example}

\subsection{Problem statement and solution overview}
The fundamental challenges encountered while developing an FL approach to mitigate the heterogeneity are: \textit{1) how to reduce the training and communication time of the stragglers in FL?, 2) how to achieve adequate performance within the fixed time interval for communication?, and 3) how to minimize the error gap between actual and inconsistent objective functions due to heterogeneous participants?} In this work, we investigate and solve the problem of \textit{training the local model on all the heterogeneous participants within a given maximum allowable response time, achieving adequate performance and minimizing error due to inconsistent objective function.} 

Apart from the standard FL workflow, the Fed-RAC trains the local models on all the participants despite higher heterogeneity and reduces training time without compromising performance. Fed-RAC starts with the estimation of the optimal number of clusters to accommodate all $N$ heterogeneous participants. We named the step as \textit{resource aware clustering} (Section~\ref{rac}). During clustering, a set $\mathcal{K}$ of $k$ clusters is first identified (Section~\ref{rac1}), followed by the generation of a generic model for each cluster (Section~\ref{rac2}). Next, the participants are assigned to the empty clusters using \textit{participant assignment mechanism} (Section~\ref{pac}). Further, we introduce \textit{master-slave technique} (Section~\ref{mspi}) to enhance the performance of the generic models using KD.

\section{Fed-RAC: \underline{Fed}erated Learning via \underline{R}esource \underline{A}ware \underline{C}lustering}\label{approach}
\begin{figure*}[h]
\centering
\includegraphics[scale=0.84]{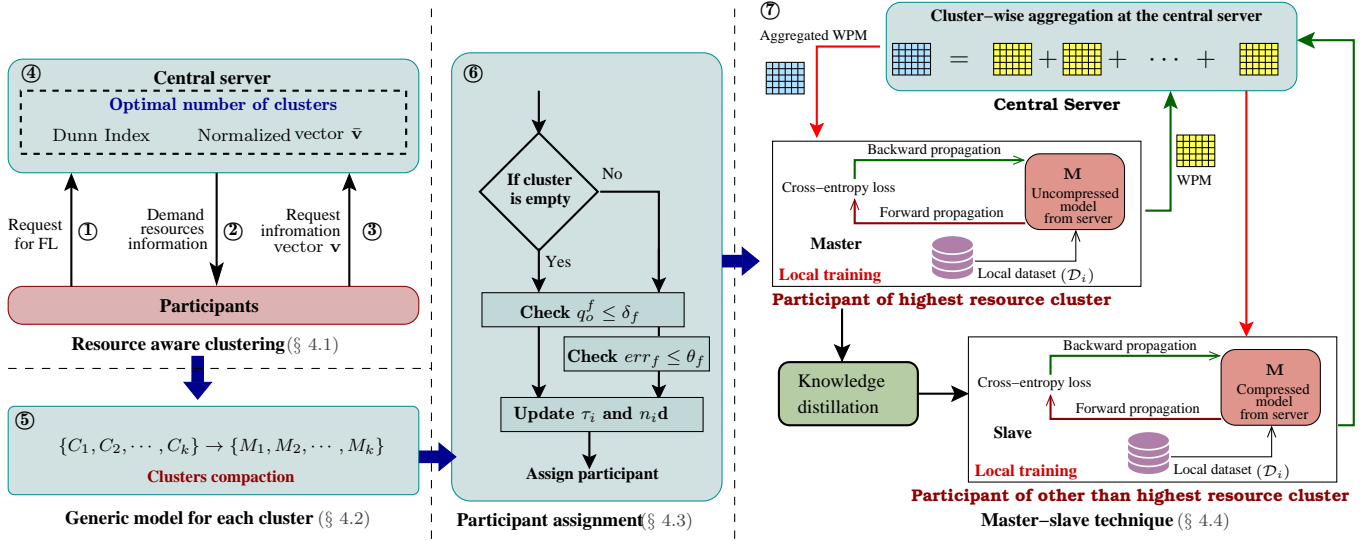}
\caption{Workflow for Fed-RAC approach. $\textcircled{1}-\textcircled{4}$ steps for resource aware clustering, $\textcircled{5}$ generating generic model for each cluster, $\textcircled{6}$ participants assignment to the clusters, and $\textcircled{7}$ master-slave technique to improve performance. }
\label{framework}
\end{figure*}

In this section, we first cover the details of the \underline{Fed}erated learning approach to mitigate the heterogeneity of participants using \underline{R}esource \underline{A}ware \underline{C}lustering (Fed-RAC). The workflow of the Fed-RAC is shown in Fig.~\ref{framework}.

\subsection{Resource aware clustering}\label{rac}
\textcolor{black}{This sub-section describes the mechanism of dividing the set of $N$ participants into $k$ disjoint clusters. The clustering is performed on the server to preserve the resources of the participants. In doing so, the server fetches three resources from all the participants, \textit{i.e.,} processing speed, data transmission rate, and available memory, denoted as $s_i$, $r_i$, and $a_i$ for $p_i$ ($1\le i \le N$), respectively. $s_i$ and $a_i$ are the machine-dependent parameters that rely upon the configuration of the devices. The data transmission rate $r_i$ depends on the bandwidth, channel coefficient, and path loss between participant and server and is estimated using technique discussed in~\cite{9066896}. The static information of $s_i$, $r_i$, and $a_i$ from the participants are used to initialized the Fed-RAC approach. Afterward, the approach provides the opportunity to upgrade or downgrade the cluster depending on the available dynamic resources of the participants. If a participant is in the smallest cluster and its resources are dynamically reduced then Fed-RAC sets batch-size and local epochs to continue the training, as discussed in Section~\ref{assignment1}. It implies the Fed-RAC can easily tackle the dynamic resources of the participants in FL.} 

All the participants of a cluster possess similar processing speed, transmission rate, and memory. However, it is tedious to determine the similarity among the three independent resources. Thus, we use a vector $v_i=[s_i,r_i,a_i]$ for participant $p_i$ ($1\le i \le N$) to estimate similarity among resources. We use normalize vector $\bar{v_i}=[\bar{s_i},\bar{r_i},\bar{a_i}]$ in place of $v_i$, to eliminate impact of biasness of high values. The bias value $\bar{s_i}$ is estimated as: $\bar{s_i}=\frac{s_i-\min\{s_i\}_{i=1}^{N}}{\max\{s_i\}_{i=1}^{N}-\min\{s_i\}_{i=1}^{N}}$, $\bar{r_i}$ and $\bar{a_i}$ are also estimated similarly. We further estimate the similarity ($\mathcal{S}_{ij}$) among any two participant $p_i$ and $p_j$ using normalized vectors $\bar{v_i}$ and $\bar{v_j}$, respectively, $\forall i, j \in\{1,2,\cdots, N\}$ using Euclidean distance. $\mathcal{S}_{ij}$ is estimated as: $\mathcal{S}_{ij}=\sqrt{\lambda_1(\bar{s_i}-\bar{s_j})^2+\lambda 
_2(\bar{r_i}-\bar{r_j})^2+\lambda_3(\bar{a_i}-\bar{a_j})^2}$, where $\lambda_1$, $\lambda_2$ and $\lambda_3$ are the contributions of processing capacity, transmission rate, and memory, respectively, $\lambda_1+\lambda_2+\lambda_3=1$. $\lambda_1$, $\lambda_2$, and $\lambda_3$ can be obtained from analysis given in~\cite{resource,fastdeepiot}. 

\subsubsection{Estimating optimal number of cluster $k$}\label{rac1} 
We introduce a modified version of the conventional Dunn and Dunn-like Indices~\cite{6170593} to estimate the optimal number of clusters using similarity. We use $k$-means clustering to determine the optimal number of clusters. Dunn index identifies an optimal number of clusters that hold compactness and provide good separation. Let $C_f$ and $C_g$ denote clusters in $\mathcal{K}$ ($C_f, C_g \in \{C_1, \cdots, C_{k}\}, C_f\ne C_g$). The least distance $\mathbf{dist}(C_f, C_g)$ among $C_f$ and $C_g$ is given as:
\begin{equation}\label{t1}
 \mathbf{dist}(C_f,C_g) = \min_{p_i \in C_f, p_j \in C_g, C_f \ne C_g } \mathcal{S}_{ij}.
\end{equation}

The diameter $\mathbf{dia}(C_f)$ of cluster $C_f\in\{C_1, \cdots, C_{k}\}$ is the maximum distance between any two participants in $C_f$. Let $p_l^f$ and $p_q^f$ be the two participants in $C_f$ ($p_l^f \ne p_q^f$), $\mathbf{dia}(C_f)$ is estimated as:
\begin{equation}\label{t2}
 \mathbf{dia}(C_f) = \max_{p_l^f, p_q^f \in C_f, p_l^f \ne p_q^f } \mathcal{S}^f_{lq}.
\end{equation}

\noindent Using Equations~\ref{t1} and~\ref{t2}, we estimate Dunn Index ($DI(k)$) as:
\begin{equation}\label{di}
 \hspace{-0.2cm} DI(k) = \min_{\forall C_f \in \mathcal{K}} \Big[\min_{\forall C_g \in \mathcal{K}, C_f \ne C_g}\Big(\frac{\mathbf{dist}(C_f,C_g)}{ \max_{\forall C_f \in \mathcal{K}} \mathbf{dia}(C_f)}\Big)\Big].
\end{equation}

A high positive value of $DI(\cdot)$ indicates compact and adequate number of clusters. The divergence-based Dunn and Dunn-like Indices starts with $k=2$ and terminates when $DI(\cdot)$ achieves a higher positive value. We use the maximum number of clusters $k_{max}\le \sqrt{N}$ as rule of thumb, inspired from~\cite{bezdek1998some}. The complete steps to obtain optimal number of clusters are given in Procedure~$1$.

\begin{example}\label{ex1}
Let there are $10$ participants denoted as $p_1, \cdots, p_{10}$. The resource and normalized vectors of the example are shown in Table~\ref{table1}. Using Procedure~$1$ with $\lambda_1=\lambda_2=\lambda_3=1/3$, we obtain $k=3$ as optimal clusters. 
\begin{table}[h]
\centering
\caption{An illustration of example scenario having $10$ participants with resource vectors and normalized resource vectors.}
\label{table1}
 \resizebox{.49\textwidth}{!}{
\begin{tabular}{|c|c|c|}
\hline
\multicolumn{1}{|c|}{\textbf{Participant}} & \multicolumn{1}{c|}{\textbf{\begin{tabular}[c]{@{}c@{}}Resource vector\end{tabular}}} & \multicolumn{1}{c|}{\textbf{\begin{tabular}[c]{@{}c@{}}Normalized resource vector\end{tabular}}} \\ \hline 
    $p_1$  &   $v_1=[100,10,20]$  &   $\bar{v_1}=[0.5,0.375,0.5]$      \\ 
    $p_2$  &   $v_2=[50,15,30]$   &   $\bar{v_2}=[0,1,1]$ \\
    $p_3$  &   $v_3=[75,8,25]$    &   $\bar{v_3}=[0.25,0.125,0.75]$ \\
    $p_4$  &   $v_4=[125,10,15]$  &   $\bar{v_4}=[0.25,0.625,0.75]$ \\
    $p_5$  &   $v_5=[150,7,10]$   &   $\bar{v_5}=[1,0,0]$ \\
    $p_6$  &   $v_6=[110,10,25]$  &   $\bar{v_6}=[0.6,0.375,0.75]$\\
    $p_7$  &   $v_7=[125,15,20]$  &   $\bar{v_7}=[0.75,1,0.5]$\\
    $p_8$  &   $v_8=[80,10,10]$   &   $\bar{v_8}=[0.30,0.375,0]$\\
    $p_9$  &   $v_9=[75,15,20]$   &   $\bar{v_9}=[0.25,1,0.5]$\\
$p_{10}$   &   $v_{10}=[50,10,30]$ &  $\bar{v_{10}}=[0,0.375,1]$ \\ \hline
\end{tabular}
}
\end{table}

\end{example}

\SetAlFnt{\small}
\begin{procedure}[t]
\caption{() \textbf{1: Optimal number of clusters}} 
\label{proc1}
\KwIn{ Set of $N$ participants $\mathcal{P}$ in FL\;}
\KwOut{Optimal set of $k$ clusters $\mathcal{K}=\{C_1, C_2, \cdots, C_{k}\}$\;} 
Initialization: $j\leftarrow0$, $C_s\leftarrow [ ]$, $\mathcal{K}=\{\}$, $k\leftarrow 2$; /*starting with $2$/*\\
\For{each participant $p_i \in \{p_1, p_2,\cdots p_N\}$}{
Server extracts information of $s_i$, $r_i$, and $a_i$ from $p_i$\;
Estimate resource vector $v_i$\;
}
\For{each participant $p_i \in \{p_1, p_2,\cdots p_N\}$}{
Estimate $\bar{s_i}$, $\bar{r_i}$, and $\bar{a_i}$ for $p_i$\;
Determine normalized resource vector $\bar{v_i}$\;
}
\While{$k \le \sqrt{N}$}{
     Perform $k$-mean clustering on set $\mathcal{P}$\;
     Estimate similarity among the normalized vectors\;
     Preserve information about the $k$ clusters\;
     \For{each pair $C_f, C_g \in \{C_1, C_2, \cdots, C_{k}\}, C_f \ne C_g $}{
         Estimate Dunn index ($DI(k)$) using \eqref{di}\;
     }
     $C_s\leftarrow$ \textit{append($DI(k)$)}, $k \leftarrow k+1$\;
} 
$j\leftarrow \arg\max(C_s)$, $k\leftarrow j+1$; /*Optimal number of clusters*/\\
\textbf{return} $\mathcal{K}=\{C_1, C_2, \cdots, C_k\}$;
\end{procedure}

\textcolor{black}{Apart from $k$-means clustering, we also consider density-based clustering to obtain the optimal number of clusters using normalized resource vectors. We use Density-Based Spatial Clustering of Applications with Noise (DBSCAN) and  Ordering points to identify the clustering structure (OPTICS)~\cite{6493211} during the experiment. Table~\ref{c1} illustrates the DI-values and accuracy for different $k$ using k-means, DBSCAN, and OPTICS using the resource vectors, discussed in Section~\ref{e01}. From the results in the table, we observe that for DBSCAN clustering, the DI value decreases with increasing $k$; thus, it predicts $k=2$ as an optimal number of the cluster. However, the difference between resources among the participants within a cluster is high, which results in lower accuracy. Moreover, some participants with the least resources can not accommodate a large-size model assigned to the cluster. We can draw similar observations for OPTICS, which gives the optimal number of clusters $k=3$. k-means clustering results in $k=5$ optimal number of clusters, where inter-cluster and intra-cluster distances are high and low, respectively. It narrows downs the gaps between the resources of the participants within a cluster. Thus, all the participants can easily accommodate the assigned model to a cluster. Such narrow gapping also prevents the bucket effect, where a large model is assigned to the participant with the smallest resources.}

\begin{table}[h]
\caption{Impact of clustering techniques on DI values and accuracy at different values of $k$ using MNIST dataset.}
\resizebox{.49\textwidth}{!}{
\begin{tabular}{|c|ccccc|}
\hline
\multirow{2}{*}{\textbf{\begin{tabular}[c]{@{}c@{}}Cluster\\ technique\end{tabular}}} & \multicolumn{5}{c|}{\textbf{DI values}}                                                                                                   \\ \cline{2-6} 
                                & $\mathbf{k=2}$    & $\mathbf{k=3}$    & $\mathbf{k=4}$ & $\mathbf{k=5}$   & $\mathbf{k=6}$ \\ \hline
k-means                         & $0.1517$          & $0.1965$          & $0.2165$       & $\mathbf{0.2317}$& $0.1750$     \\ \hline
DBSCAN                          & $\mathbf{0.2231}$ & $0.1819$          & $0.1642$       & $0.1419$         & $0.1236$         \\ \hline
OPTICS                          & $0.1165$          & $\mathbf{0.1208}$ & $0.1037$       & $0.0839$         & $0.0673$       \\ \hline \hline
\textbf{Accuracy ($\pm0.30$)}   & $94.39\%$         & $95.07\%$         & $96.32\%$      & $\mathbf{97.73\%}$        & $95.67\%$       \\ \hline
\end{tabular}
}
\label{c1}
\end{table}

\subsubsection{Generic model for each cluster and compaction of clusters}\label{rac2} 
This work considers three resources, \textit{i.e.,} processing speed, data transmission rate, and available memory, to obtain $k$ clusters. However, the cumulative resources are unequal among all the clusters. Therefore, the size of the model on the clusters would be non-identical in FL. This work develops a generic model for each cluster and performs cluster compaction afterward. In doing so, we arrange the $k$ clusters in descending order of their available resources. In other words, the participants in cluster $C_1$ can train a large-size model and quickly transfer WPM to the server, whereas  $C_k$ can train the smallest model and requires the huge time to share WPM. 

Let $M$ denotes the initial model generated and randomly initialized by the server. We assume $M$ can directly accommodate on $C_1$, \textit{i.e.,} training and communication can be performed within the given time. Let $M_1$ denotes the size of model for $C_1$, where $M=M_1$. Beyond $C_1$ other clusters require some compression to train the model and share WPM. Let $M_2$ denotes the compressed version of $M$ that can be deployed on the participants in $C_2$, consuming less training and communication time. Similarly, $M_3-M_k$ are generated for remaining $k-2$ clusters. In this work, we consider the model of any cluster $C_i$ is $\alpha$ times smaller than $C_{i-1}$, \textit{i.e.,} $M_{i-1}=\alpha M_i$, where $\alpha<1$. It implies $M_k=\alpha^{k-1}M_1 \implies M_k=\alpha^{k-1}M$.

\noindent $\bullet$ \textbf{Cluster compaction:} The estimated $k$ clusters and corresponding models suit the resources of the participants; however, higher compression of the model results in performance compromise. Thus, it is beneficial if all the participants can accommodate in fewer clusters than $k$. However, it introduces the \textit{straggler effect}, where slow participants do not participate. To overcome the straggler effect, we merge some clusters out of $k$ to obtain $m$ clusters, where $k<m$. 

\subsection{Participants assignment to the clusters}\label{pac}
This sub-section describes the mechanism of assigning $N$ participants to the $m$ clusters. We first deduce the expression to estimate the communication rounds required for the generic model in $m$ different clusters. Next, we define the optimization error due to the heterogeneity of participants. Notably, Fed-RAC initially checks the possibilities of assigning participants to the higher cluster, decreasing as per the assignment criteria.

From Section~\ref{rac2}, we have $m$ different models $M_1, M_2,\cdots,M_m$ for clusters $C_1, C_2, \cdots, C_m$, respectively, where the size of models $M_1>M_2>\cdots>M_m$ and $M_m=\alpha^{m-1}M_1=\alpha^{m-1}M$. The server decides $\mathcal{R}_1, \mathcal{R}_2, \cdots, \mathcal{R}_m$ communication rounds for training local models of the participants in clusters $C_1, C_2, \cdots, C_m$, respectively. We first determine the expression for communication rounds $\mathcal{R}_f$  for cluster $C_f$, where $1\le f\le m$. 

\subsubsection{Communication rounds per cluster}
Let $\mathcal{P}_f$ denotes the set of $F$ participants to be assigned in $C_f$, where $\mathcal{P}_f=\{p_1,\cdots, p_F\}$, having loss functions $\mathcal{L}_1, \cdots   \mathcal{L}_F$, respectively. We consider the assumptions given in~\cite{li2019convergence,mishra2022suppressing} and applied on  $\mathcal{L}_1, \cdots, \mathcal{L}_F$ to estimate the round $\mathcal{R}_f$ for cluster $C_f$. 
\begin{assumption}\label{a1}
Loss $\mathcal{L}_j \in \{\mathcal{L}_1, \mathcal{L}_2, \cdots   \mathcal{L}_F\}$ is $L$-smooth; therefore, for any two WPM $\mathbf{w}_a$ and $\mathbf{w}_b$ on $p_j\in \mathcal{P}_f$, following inequality holds: $\mathcal{L}_j(\mathbf{w}_a)\leq \mathcal{L}_j(\mathbf{w}_b) + (\mathbf{w}_a-\mathbf{w}_b)^T\nabla \mathcal{L}_j(\mathbf{w}_b)+ \frac{L}{2}\|\mathbf{w}_a-\mathbf{w}_b\|^2$, where $1\le j \le F$. 
\end{assumption}

\begin{assumption}\label{a2}
 $\mathcal{L}_j$ is $\mu$-strongly convex; the inequality holds: $\mathcal{L}_j(\mathbf{w}_a)\geq \mathcal{L}_j(\mathbf{w}_b) + (\mathbf{w}_a-\mathbf{w}_b)^T\nabla \mathcal{L}_j(\mathbf{w}_b)+ \frac{\mu}{2}\|\mathbf{w}_a-\mathbf{w}_b\|^2$.  
\end{assumption}

\begin{assumption}\label{a3}
Let $\varepsilon_j^{t}$ denotes the uniformly and randomly selected sample from the local dataset $\mathcal{D}_j$ of participant $p_j$ on communication round $t$, where $1\le t \le \mathcal{R}_f$. Let $\nabla\mathcal{L}_j(\varepsilon_j^{t}, \mathbf{w}_j^{t})$ and $\nabla\mathcal{L}_j(\mathbf{w}_j^{t})$ denote the gradients of loss function $\mathcal{L}_j(\cdot)$ on $\varepsilon_j^{t}$ samples and entire samples of the local dataset, respectively. The variance of gradients on participant $p_j$ is bounded as: $\mathbb{E}\|\nabla\mathcal{L}_j(\varepsilon_j^{t}, \mathbf{w}_j^{t})-\nabla\mathcal{L}_j(\mathbf{w}_j^{t})\|^2 \leq \sigma_f^2$. 
\end{assumption}

\begin{assumption}\label{a4}
Expected square norm of loss gradient is uniformly bounded as: $\mathbb{E}\|\nabla\mathcal{L}_j(\phi_j^{t}, \mathbf{w}_j^{t})\|^2 \leq G^2_f$, $1\le {t} \le \mathcal{R}_f$ and $1\le j \le F$. 
\end{assumption}

Using Assumptions~$1-4$, we obtain a relation between desired precision $(q_o^f)$, local epoch count $E_f$, and global iterations $\mathcal{R}^f$ of cluster $C_f$. The precision is defined as: $q_o^f=\mathbb{E}[\mathcal{L}(\mathbf{w}^{\mathcal{R}_f})]-\mathcal{L}_f^{*}$, where $\mathbf{w}^{\mathcal{R}_f}$ is the aggregated weight at final global epoch $\mathcal{R}_f$ and $\mathcal{L}_f^{*}$ is minimum and unknown  value of $\mathcal{L}_f$ at the server. Let $\mathcal{L}^{*}_j$ is the minimum value of $\mathcal{L}_j$ at $p_j$, where $\forall j \in \{1\le j \le F\}$. In this work, we assume i.i.d datasets on the participants; thus, $\Gamma=\mathcal{L}_f^{*}-\sum_{i=1}^{F}\mathcal{L}_j^{*}=0$, as given~\cite{li2019convergence}. $\Gamma$ quantifies the degree of non-i.i.d and it goes to zero for i.i.d. Let $\epsilon_j$ denotes the weight contribution of participant $p_j \in \mathcal{P}_f$. Let $\beta=\max\{8L/\mu,E_f\}$ and $T_f$ is the total SGD operations on a participant then we obtain following relation of desired precision ($q_o^f$) for cluster $C_f$~\cite{li2019convergence,mishra2022designing}:
\begin{align}\small\label{e9}
\mathbb{E}[\mathcal{L}(\mathbf{w}^{\mathcal{R}_f})]-\mathcal{L}^{*}_f  \le \frac{L/2\mu^2}{\beta+T_f-1} \Big(4B+\mu^2\beta\mathbb{E}\|\cdot\|^2\Big),
\end{align}
where, $B =  \sum_{j=1}^{F} \epsilon_j^2 \sigma_f^2 + 8(E-1)^2 G_f^2$. Using upper bound of $q_o^f$ and $T_f=\mathcal{R}_fE_f$, we obtain number of communication round ($\mathcal{R}_f$) for cluster $C_f$ ($1\le f \le m$) as follows: 
 \begin{equation}\label{e11}
 \mathcal{R}_f=\frac{1}{E_f}\Big[\frac{L}{2\mu^2 q_o^f}\Big(4B+\mu^2\beta\mathbb{E}\|\mathbf{w}_1-\mathbf{w}^{*}_f\|^2\Big)+1-\beta\Big].
\end{equation}

From \eqref{e11}, we have fixed communication rounds $\mathcal{R}^f$ for given precision threshold $q_o^f$ and local epochs $E_f$ for cluster $C_f$, where $1 \le f \le m$. In addition, we have $E_f=\frac{B_j \tau_j}{n_j}$; it implies we can change value of $B_j$, $\tau_j$, and $n_j$ in such a manner, where $E_f$ and $R_f$ remains fixed for $p_j \in \mathcal{P}_f$ and $q_o^f$ changes. We set a threshold over $q_o^f$, denoted as $\delta_f$ for $C_f$. 

\begin{example}
Let $\mu=0.7$, $L=1.5$, $B=1$, $\mathbb{E}\|\mathbf{w}_1-\mathbf{w}^{*}_f\|=0.08$ and $E_f=20$ for cluster $C_f$. We obtain $\beta=\max\{8\times1.5/0.7,20\}=20$ using $E_f$, $\mu$, and $L$. Further, we estimate $\mathcal{R}_f=6$ using \eqref{e11} and values of parameters given above. 
\end{example}

\subsubsection{Optimization error due to participants heterogeneity}
The set of participants $\mathcal{P}_f$ to be assigned in cluster $C_f$ posses low inter-cluster and high intra-cluster heterogeneity. Therefore, we obtain inconsistency in the objective function of a cluster, discussed in Section~\ref{obj}, due to intra-cluster heterogeneity despite using an effective clustering mechanism. To estimate the value of error $err_f$ for cluster $C_f$, where $C_f\in\{C_1,C_2,\cdots,C_m\}$, we use the assumptions given in~\cite{objective}. The previous assumptions, \textit{i.e.,} Assumption~\ref{a1} and Assumption~\ref{a2} are same for estimating $err_f$. However, we need to define a new assumption (Assumption~\ref{a5}) to calculate $err_f$.  

\begin{assumption}\label{a5}
Let $\{\epsilon_1,\epsilon_2,\cdots,\epsilon_F\}$ denote a set of weighted contribution of participants in set $\mathcal{P}_f$ of cluster $C_f$, where $\sum_{j=1}^{F}\epsilon_j=1$ and $C_f\in\{C_1,\cdots,C_m\}$. There exists two constants $h_1 \ge 1$ and $h_2 \ge 0$ such that $\sum_{j=1}^{F}\epsilon_j\|\nabla\mathcal{L}_j(\mathbf{w}_j)\|^2 \le h_1^2 \|\sum_{j=1}^{F}\epsilon_j\nabla\mathcal{L}_j(\mathbf{w}_j)\|^2+h_2^2.$
\end{assumption}

Using Assumptions~\ref{a1}, \ref{a2}, and \ref{a5}, we derive the expression for $err_f$ of cluster $C_f$. In doing so, let $\mathbf{o}_j$ denotes a non-negative vector and defines how stochastic gradients are locally accumulated. For example, $\mathbf{o}_j=[1,\cdots,1]\in\mathbb{R}^{\tau_j}$ for FedAvg~\cite{mcmahan2017,9798299}. $\|\mathbf{o}_j\|_1$ is the $l1$-norm of $\mathbf{o}_j$ and $o_{[j,-1]}$ is the last element in vector $\mathbf{o}_j$. 
$\tau_e=\sum_{j=1}^{F}\tau_j/F$, $\tau_j=\lfloor E_f n_j /B_j\rfloor$ and $\eta$ is the learning rate, where $1\le j \le F$. 
\begin{align}\small\nonumber
 err_f & =\min_{t\in \mathcal{R}_f}\mathbb{E}\Big[\|\nabla \bar{\mathcal{L}}(\bar{\mathbf{w}}^t)\|^2\Big] \\ \label{error}
 \le & \frac{4b_1}{\eta \tau_e \mathcal{R}_f} + \frac{4 \eta L \sigma_f^2b_2}{F} + 6 \eta^2 L^2 \sigma_f^2 b_3 + 12 \eta^2 L^2 h^2_2 b_4,
 \end{align}
where $b_1=[\bar{\mathcal{L}}(\bar{\mathbf{w}}^{0})-\mathcal{L}^{*}_f], b_2 = F \tau_e \sum_{j=1}^{F}\frac{\epsilon_j^2\|\mathbf{o}_j\|_2^2}{\|\mathbf{o}_j\|_1^2},\\ b_3 = \sum_{j=1}^{F} \epsilon_j (\|\mathbf{o}_j\|_2^2-[o_{j,-1}]^2), b_4 = \max_{j} \{\| \mathbf{o}_j\|_1(\|\mathbf{o}_j\|_1-[o_{j,-1}])\}$. A small $err_f$ indicates lower intra-heterogeneity among the participants. We set error bound for each cluster, \textit{i.e.,} error $err_f\le \theta_f$ for $C_f$, where $1\le f \le m$ and $err_f\le \theta_f$.

\subsubsection{Participants assignment}\label{assignment1} 
Fed-RAC assigns each participant to an optimal cluster per the available device and networking resources. Such assignment facilitates easier and faster (within MAR time) training and inference of the local model on each participant assigned to a specific cluster. In other words, each participant trains the local model in $R_f$ communication rounds (\eqref{e11}) for cluster $C_f$, $1\le f \le m$. The assignment verifies two conditions: a) precision \eqref{e9} of cluster $C_f$ must be less than the threshold ($q_o^f\le \delta_f$) and b) optimization error \eqref{error} $err_f \le \theta_f$. Further, we get two possible cases for assigning participants in each cluster:

\noindent $\bullet$ \textbf{Case 1 (Cluster is empty):} $p_i$ assigns to empty cluster $C_f$, if $p_i$ can train the model $M_f$ in given epochs $E_f$ and communication round $R_f$. The local epoch $E_f$ is high for a single participant as one communication round is required to train the model without multiple participants. In this case, the condition of $q_o^f<\delta_f$ is only verified and the optimization error is zero. It is because the constraint for homogeneity becomes zero with a single participant in \eqref{error}. If the participant is unable to train $M_f$ in MAR and $R_f$, it uses the following two steps:
\begin{enumerate}
\item $p_i$ reduces $\tau_i$ and $n_i$, while satisfying $q_o^f\le \delta_f$.
\item If $q_o^f\ge \delta_f$ for $C_f$ then the participant switches to the lower cluster and repeats Step 1.
\end{enumerate}

\noindent $\bullet$ \textbf{Case 2 (Cluster is non-empty):}
We initially estimate  $q_o^f$ \eqref{e9}. Upon adding $p_i$ to $C_f$, $q_o^f$ should be less than threshold $\delta_f$. Similar to Case 1, if $p_i$ is incompetent in training $M_f$ in MAR, $\tau_i$ and $n_i$ are adjusted until $q_o^f< \delta_f$; otherwise participant switches to the lower cluster. Next, the error \eqref{error} is also estimated upon adding $p_i$ to $C_f$. If estimated $err_f \le \theta_f$ then $p_i$ is added to $C_f$, else $p_i$ switches to lower cluster. 

\textcolor{black}{After successfully executing these two cases, $N$ participants are assigned to the $m$ clusters. The assigned participants achieve desired precision and optimization errors less than the corresponding thresholds. The server optimally allocates each participant to a specific cluster as per the resource, precision threshold, and error threshold. Procedure~$2$ summarizes the steps involved in assigning participants to the clusters. }

\SetAlFnt{\small}
\begin{procedure}[t]
\caption{() \textbf{2: Participants assignment to the clusters}} 
\label{proc1}
\KwIn{Set of clusters $\{C_1,\cdots, C_m\}$ with generic models $\{M_1,\cdots, M_m\}$. Set of participants $\mathcal{P}=\{p_1,\cdots, p_N\}$\;}
\KwOut{Optimal participants in each cluster $C_f$, $\forall 1\le f \le m $\;} 
Initialization: $i\leftarrow 1$, $f\leftarrow1$\;
\For{each participant $p_i \in \{p_1, p_2,\cdots p_N\}$}{
\For{each cluster $C_f \in \{C_1,C_2,\cdots, C_m\}$}
    {
        \nonl /*\textit{Case 1 for assigning participant}*/\\
        \If{$isEmpty(C_f)==True$}{
            \If{$p_i$ can accommodate $M_f$}
            {
                \textit{Check:} Estimate precision $q_o^f$ using \eqref{e9}\;
                \If{$q_o^f \le \delta_f$}{
                Assign $p_i$ to $C_f$\;
                }
                \Else{
                    $f\leftarrow f+1$\;}
                }
            \Else{
                Reduce $\tau_i$ and $n_i$ \textit{s.t.}, $p_i$ can run $M_f$\;
                \textbf{Goto} \textit{Check}\;
                }
            }
        \nonl /*\textit{Case 2 for assigning participant}*/\\
        \Else{
            \If{$p_i$ can accommodate $M_f$}
            {
                \textit{Check-I:} Estimate precision $q_o^f$ using \eqref{e9}\;
                Calculate error $err_f$ using \eqref{error}\;
                \If{$q_o^f \le \delta_f$ \text{and} $err_f \le \theta_f$}{
                    Assign $p_i$ to $C_f$\;
                    }
                \Else{
                    $f\leftarrow f+1$\;}
            }
            \Else{
                Reduce $\tau_i$ and $n_i$ \textit{s.t.}, $p_i$ can run $M_f$\;
                \textbf{Goto} \textit{Check-I}\;
                }
            }

    }
}
\textbf{return} Optimal participants in each cluster $C_f$, $\forall 1\le f \le m $\;
\end{procedure}

\subsection{Master-slave technique}\label{mspi}
This sub-section introduces the technique of improving the performance of lightweight models $M_2, \cdots,M_m$ in clusters $\{C_2, \cdots,C_m\}$ using generalization ability (or knowledge) of large-size model $M_1$ in cluster $C_1$. We utilize the assumption that the cluster $C_1$ is the fastest cluster and can accommodate the server's model without compression, \textit{i.e.,} $M_1=M$. We use the term \textit{master} for $M_1$ and \textit{slave} for models $M_2-M_m$, thus, named the technique as \textit{master-slave for performance improvement}. The technique involve KD technique~\cite{hinton2015distilling,10.1145/3570955,9994660} to improve the performance of slave model using trained master model. MAR time ($\mathbb{T}_{max}$) for training models on all $N$ participants and can be further divided as: $\mathbb{T}_{max}=T_1+\max\{T_2,T_3,\cdots,T_m\}$, where $T_f$ is the MAR time for training $M_f$ on the participants of $C_f$, $1\le f \le m$. Since $C_m$ is the slowest cluster and $C_1$ is the fastest cluster; thus, we can consider the following relation similar to generic models: $T_{f-1}=\kappa T_f$, where $1\le f \le m$ and $\kappa<1$. It implies $T_1=\kappa^{m-1}T_m$  then we obtain:
\begin{align}\nonumber
 \mathbb{T}_{max}&=\kappa^{m-1}T_m + \max\{\kappa^{m-2}T_m, \kappa^{m-3}T_m,\cdots, T_m\}\},\\
 &=\kappa^{m-1}T_m+T_m = (\kappa^{m-1}+1)T_m.
\end{align}

In special case, where $M_1$ is master of $M_2$, $M_2$ is master of $M_3$, and so on, \textit{i.e.,} the FL-based training is performed sequentially for each cluster. In this case, $\mathbb{T}_{max}$ is defined as:
\begin{align}\nonumber
 \mathbb{T}_{max}&=\kappa^{m-1}T_m + \kappa^{m-2}T_m + \kappa^{m-3}T_m + \cdots + T_m,\\ \nonumber
 =\{\kappa^{m-1}  &+   \kappa^{m-3} + \cdots + 1\}T_m =\frac{1-\kappa^{m}}{1-\kappa}, \text{ where } \kappa <1.
\end{align}
 
This work starts FL based training from the fastest cluster $C_1$ with the adequate devices and networking resources to train $M_1$. We train $M_1$ for $E_1$ local epochs on the participants of $C_1$ using $\mathcal{R}_1$ communication rounds. The logits of trained $M_1$ is next supplied to all the remaining clusters to improve the performance of their generic models using the KD. Algorithm~\ref{algo1} summarizes steps involved in the Fed-RAC. 

Fed-RAC approach trains the lightweight models of the smaller clusters using the knowledge distillation (master-slave) technique. It may lead to biased learning because knowledge extracted from the data samples in the larger cluster is utilized more times than those in smaller ones. To avoid such biaseness, we adopt the resampling and reweighting scheme. It resolves the trade-off between minor and frequently chosen data instances for model training on the larger cluster. In other words, the participants of the largest cluster sample nearly equal number of data instances for all the classes during training in each communication round.

\SetAlFnt{\small}
\begin{algorithm}[t]
\caption{\textbf{Fed-RAC Algorithm.}}
\label{algo1}
\KwIn{Set $\mathcal{P}$ of $N$ participants with their local datasets;}
Call \textbf{Procedure~1} to determine a set $\mathcal{K}$ of $k$ clusters\;
Merge clusters to obtain $m$ clusters in $\mathcal{K}$, $\{C_1,C_2,\cdots,C_m\}$\;
Generate $m$ model for each cluster, $\{M_1,M_2,\cdots, M_m\}$\;
/*Participants assignment to the clusters*/\\
Call \textbf{Procedure~2} to assign optimal number of participants to all the $m$ clusters\;
\For{each cluster $C_f \in \{C_1,C_2,\cdots, C_m\}$}
    {
        \If{$f==1$}{
          \While{communication rounds $ r \le R_1$ }{
            Train local models in cluster $C_1$\;
            $r\leftarrow r+1$\;
          }
      Obtained train model $M_1$ for participants in cluster $C_1$\;
      }
      \Else
      {
            \While{communication rounds $ r \le R_f$ }{

            Train local models in cluster $C_f$ under the guidance of model $M_1$\;
            $r\leftarrow r+1$\;
           } 
       Obtained train model $M_f$ for participants in cluster $C_f$\;    
      }
 }
\textbf{return} Trained model on each participant\; 
\end{algorithm}

\section{Performance Evaluation}\label{evaluation}
This section describes the task of study, datasets, and models, followed by the baselines. We consider the different tasks, including Locomotion Mode Recognition (LMR),  Human Activity Recognition (HAR), Handwritten Digit Recognition (HDR), and Image Classification (IC).

\subsection{Datasets and models}
This work uses four public datasets, including MNIST~\cite{lecun1998gradient}, HAR~\cite{anguita2013public}, CIFAR-10~\cite{krizhevsky2009learning}, and SHL~\cite{shl2}. These datasets were selected due to free accessibility, real-life acquisition, and correct annotations. MNIST is a handwritten digit dataset containing $50000$ images of different digits from $0-9$ for training. MNIST also has $10000$ images for testing. HAR was collected using the smartphone (Samsung Galaxy S II) sensors, including a tri-axial accelerometer and gyroscope. The dataset contains sensory instances of six different activities; walking, standing, lying, sitting, walking upstairs, and walking downstairs. CIFAR-10 comprises $60000$ images of ten different classes. The dataset is balanced and correctly annotated with $6000$ images for each class and contained $50000$ images for training and $10000$ for testing. SHL~\cite{shl2} dataset was collected from the onboard sensors of HUAWEI Mate 9 smartphones to recognize the locomotion modes of the users.

We use a simplified arrangement of Convolutional Neural Networks (CNN) and Dense layers to obtain a model as:  $C(128)-C(64)-C(128)-C(256)-C(512)-D(classes\_count)$, where $C(X)$ indicates the convolutional layer with $X$ filter units and $D(Y)$ is the dense layer having $Y$ neurons. We use ResNet-18~\cite{he2016deep} and DeepZero~\cite{9164991,9653808} for training on CIFAR-10  and the SHL datasets, respectively. 

\subsection{Baselines}
We considered the existing techniques~\cite{diao2020heterofl, li2020federated, mcmahan2017communication, lai2021oort} as baselines, noted as HeteroFL~\cite{diao2020heterofl}, FedProx~\cite{li2020federated}, FedAvg~\cite{mcmahan2017communication}, and Oort~\cite{lai2021oort}, to evaluate and compare the performance. {HeteroFL~\cite{diao2020heterofl}} partitioned the heterogeneous participants into various clusters depending on the different computational complexities. {FedProx~\cite{li2020federated} handled the problem of heterogeneity by introducing a proximal term. The term minimized the impact of local updates and restricted such updates close to global model. {FedAvg~\cite{mcmahan2017communication} is the benchmark and most classical FL learning technique. {Oort~\cite{lai2021oort} selected a set of participants that achieved adequate accuracy and quickly trained the model. 
It is a participant selection mechanism that performed cherry-picking of the participants.    
%\end{itemize}

\subsection{System implementation}\label{implementation}
We implemented the Fed-RAC algorithm and procedures using Python programming language. The considered models are implemented using the functional API of Keras in Python language to friendly support developers. Additionally, we reimplemented all the baselines to perform a fair comparison. During the experiment, we set the loss function to ``categorical cross-entropy'', batch size to $200$ and other settings as discussed in~\cite{9479778,mishra2019fault,10041779}. We set $\mathcal{L}^{*}$ between $0.01-0.05$ and the number of participants $N=40$. The local epochs vary over the dataset, \textit{i.e.,} $E=1-5$ for MNIST and HAR, while $E=10-40$ for CIFAR-10 and SHL. The communication rounds were set to $200$ for all the clusters during the experiment. We varied the learning rate between $0.001$ to $0.010$ and set the proximal term in FedProx to $0.001$. We only compress the convolutional layers to obtain slave models. We use the dropout of $0.5$, \textit{i.e.,} $M_2=0.5(M_1)$, $M_3=0.5(M_2)$, \textit{etc.} inspired from~\cite{salehinejad2019ising,9768100,9130098}.

\subsection{Evaluation strategy}
The primary motive of FL is to improve the local performance and generalization ability. We adopt these strategies: \\
%\begin{enumerate}
\noindent \textbf{(1) Local performance:} It determines: \textit{how well the local model is trained on the dataset of the participants?}\\ 
\noindent \textbf{(2) Cluster performance:} It estimates: \textit{how well the participants can improve the cluster-wise performance through the aggregation of WPM?}\\ 
\noindent \textbf{(3) Global performance:} It is the simple average over cluster performance and helps to determine: \textit{how much deviation is observed in the cluster performance from the average value?}
%\end{enumerate}

\subsection{Evaluation metrics}
We use the standard metrics, including \textbf{accuracy} and \textbf{F1-score}, to evaluate the performance of the Fed-RAC. We also introduce a new performance metric, namely \textbf{rounds-to-reach x\%}." Let $I(x\%)$ denotes the symbolic representation of the metric. $I(x\%)$ counts the number of iterations (or rounds) required for achieving the performance of $x\%$. We finally use\textbf{leave-one-out-test} metric that trains the model for all class labels except for one randomly chosen class label. . 

\begin{table}[h]
\centering
\caption{Available resources set $\mathcal{P}$ of $40$ participants. R. Vector $=$ Resource vector $=$ [processing,
transmission rate, and memory].}
\label{tab3}
\resizebox{.50\textwidth}{!}{
\begin{tabular}{|l|@{}c@{}|l|@{}c@{}|l|@{}c@{}|l|@{}c@{}|}
\hline
$\mathcal{P}$ & \textit{R. Vector} & $\mathcal{P}$ & \textit{R. Vector} & $\mathcal{P}$ & \textit{R. Vector}& $\mathcal{P}$ & \textit{R. Vector}\\ \hline \hline
$p_1$ & $[1.6,10.88,8]$ & $p_{11}$ & $[1.6,12.54,6]$ & $p_{21}$ & $[1.6,40,1]$ & $p_{31}$ & $[3.1,18.04,6]$ \\ 
$p_2$ & $[2.8,4.1,3]$   & $p_{12}$ & $[0.8,1.2,6]$   & $p_{22}$ & $[1.1,11.4,6]$ & $p_{32}$ & $[2.5,44.13,6]$                    \\ 
$p_3$ & $[1.1,1.13,6]$  & $p_{13}$ & $[1.3,28.41,6]$ & $p_{23}$ & $[2.5,25,6]$  & $p_{33}$ & $[2.3,6.5,6]$                 \\ 
$p_4$ & $[1.6,11.45,3]$ & $p_{14}$ & $[1.3,21.9,3]$  & $p_{24}$ & $[2.2,30,4]$  & $p_{34}$ & $[2.1,60.21,6]$               \\ 
$p_5$ & $[3.2,8.9,3]$   & $p_{15}$ & $[3.1,25.99,6]$ & $p_{25}$ & $[1.6,9.62,6]$ & $p_{35}$ & $[2.1,61.3,8]$              \\ 
$p_6$ & $[2.2,2,4]$     & $p_{66}$ & $[3.2,19.43,4]$ & $p_{26}$ & $[2.2,23.27,6]$ & $p_{36}$ & $[3.2,19,6]$              \\ 
$p_7$ & $[3.1,8.7,1]$   & $p_{17}$ & $[1.0,20.98,3]$ & $p_{27}$ & $[1.5,49.79,6]$ & $p_{37}$ & $[2.7,32.05,6]$              \\ 
$p_8$ & $[1.8,60,3]$   & $p_{18}$ &  $[1.6,30,3]$    & $p_{28}$ & $[1.7,37.65,6]$ & $p_{38}$ & $[2.9,6.52,6]$              \\
$p_9$ & $[2.7,8.89,3]$   & $p_{19}$ & $[1.0,12,2]$   & $p_{29}$ & $[3.1,15.71,6]$  & $p_{39}$&  $[0.8,38.8,6]$             \\
$p_{10}$ & $[1.4,34.5,8]$ & $p_{20}$ & $[2.7,10,6]$  & $p_{30}$ & $[2.6,3,6]$      & $p_{40}$ &  $[2.1,32,6]$        \\\hline
\end{tabular}
}

\end{table}

\begin{table*}[h]
\centering
\caption{Impact of resource aware clustering on considered MNIST, HAR, CIFAR-10, and SHL datasets using different types of resource vectors. The reported results depicts the global accuracy and F1-score achieved by the different models.}
\label{tab31}
\resizebox{1.00\textwidth}{!}{
\begin{tabular}{|@{}c@{}|@{}c@{}|c|c|c|c|c|c|c|c|}
\hline
\multicolumn{1}{|c|}{\multirow{3}{*}{\textbf{\begin{tabular}[c]{@{}c@{}}Types of \\ resource vectors\end{tabular}}}} & \multicolumn{1}{c|}{\multirow{3}{*}{\textbf{\begin{tabular}[c]{@{}c@{}}Number of\\ clusters\end{tabular}}}} & \multicolumn{8}{c|}{\textbf{Datasets}}                                                                                                                                                                                                                                                                           \\ \cline{3-10} 
&    & \multicolumn{2}{c|}{\textbf{MNIST}}  & \multicolumn{2}{c|}{\textbf{HAR}}  & \multicolumn{2}{c|}{\textbf{CIFAR-10}}                                          & \multicolumn{2}{c|}{\textbf{SHL}}                          \\ \cline{3-10} 

&   & \multicolumn{1}{c|}{\textbf{Accuracy}} & \multicolumn{1}{c|}{\textbf{F1-score}} & \multicolumn{1}{c|}{\textbf{Accuracy}} & \multicolumn{1}{c|}{\textbf{F1-score}} & \multicolumn{1}{c|}{\textbf{Accuracy}} & \multicolumn{1}{c|}{\textbf{F1-score}} & \multicolumn{1}{c|}{\textbf{Accuracy}} & \textbf{F1-score} \\ \hline

\textbf{Unnormalized}  & $\mathcal{K}=4$ & $97.13\pm0.30$ & $98.06\pm 0.25$  &  $91.56 \pm 0.45 $     & $92.47 \pm 0.35$    & $90.12\pm 0.60$   & $90.83\pm 0.30$           & $89.23 \pm 0.30$   & $90.41\pm 0.40$                  \\ \hline

\begin{tabular}[c]{@{}c@{}}\textbf{Normalized}\\ \{$\lambda_1=\lambda_2=\lambda_3=1/3$\}\end{tabular}  &   $\mathcal{K}=6$                                                & $97.41\pm 0.25$ & $98.19 \pm 0.25$ &    $92.46\pm 0.40$  & $93.38 \pm 0.30 $   & $90.67\pm 0.45$   & $91.41\pm 0.30$  & $89.59 \pm 0.25$  & $90.83 \pm 0.40$                   \\ \hline

\begin{tabular}[c]{@{}c@{}}\textbf{Normalized}\\ \{$\lambda_1=\lambda_2=0.4,\lambda_3=0.2$\}\end{tabular}    &  $\mathcal{K}=5$                                                                                                              &  $97.73\pm 0.40$ & $98.37\pm 0.30$  &   $93.54\pm0.50$  &    $94.26 \pm 0.40$   & $91.01\pm 0.30$  & $92.13 \pm 0.40$   &   $90.27 \pm 0.30$ & $91.19 \pm 0.40$                  \\ \hline
\end{tabular}
}
\end{table*}

\subsection{Results}

\subsubsection{Impact of resource aware clustering}\label{e01}
This experiment aims to assess the efficacy of resource-aware clustering. The resource vectors of the devices used in the experiment are shown in Table~\ref{tab3}. The resource vector comprises processing capacity, transmission rate, and memory, and is obtained from a survey conducted on $128$ smartphone users, with prior permission obtained from the relevant authorities. From this survey, we randomly select $40$ users to create different clusters using the Fed-RAC approach, as discussed in Section~\ref{rac}. Communication rounds are set to $200$, and other parameters are described in Section~\ref{implementation}. The effectiveness of resource-aware clustering is evaluated using three types of resource vectors. The first type uses unnormalized resource vectors of the participants, whereas the second type uses normalized vectors with $\lambda_1=\lambda_2=\lambda_3=1/3$. The third type is similar to the second, but with $\lambda_1=0.4$, $\lambda_2=0.4$, and $\lambda_3=0.2$. 

Table~\ref{tab31} presents the results of evaluating the impact of normalizing resource vectors on estimating the optimal number of clusters. The findings show that un-normalized vectors yield a limited number of clusters, namely $4$ $(C_1-C_4)$, using Dunn Indices. This is due to the dominance of the transmission rate resource over other resources, resulting in non-optimal clusters. By applying unit-based normalization, all resource values are scaled into the range of $[0,1]$. The normalized values generate an optimal number of clusters using Dunn Indices, as it removes resource biasness. We obtained $6$ clusters $(C_1-C_6)$ by assigning equal contributions of all resources, \textit{i.e.,} $\lambda_1$ (processing capacity) = $\lambda_2$ (transmission rate) = $\lambda_3$ (memory) = $1/3$. When we set the contribution based on the analysis given in~\cite{fastdeepiot}, $\lambda_1=0.4$, $\lambda_2=0.4$, and $\lambda_3=0.2$, we obtained $5$ clusters ($C_1-C_5$).  

Table~\ref{tab31} presents the performance achieved by the Fed-RAC approach using different types of resource vectors on MNIST, HAR, CIFAR-10, and SHL datasets. The results show that normalizing the resource vector leads to improved performance compared to using unnormalized vectors. The normalization process is essential because when using unnormalized vectors, clustering relies on the dominating resource, leading to non-optimal clusters. These clusters may contain participants with non-identical resources that converge at irregular intervals, resulting in reduced cluster performance. Moreover, when the contributions of processing capacity ($\lambda_1$) and transmission rate ($\lambda_2$) are greater than memory ($\lambda_3$), \textit{i.e.,} $\lambda_1=\lambda_2=0.4>\lambda_3=0.2$, the cluster performance is high.

\noindent \textit{Observation: The first observation is that the normalization of the resource vector is essential to determine the optimal number of clusters. We next observed that processing capacity and transmission rate are more crucial than available memory while estimating the optimal number of clusters.}

\subsubsection{Impact of clusters compaction}\label{e02}
Table~\ref{result2} illustrates the impact of cluster compaction on the performance of Fed-RAC using MNIST, HAR, CIFAR-10, and SHL datasets. Table~\ref{result2}(a) demonstrates the cluster accuracy when all five clusters, estimated in Section~\ref{e01}, are available. The results depicted that the slave clusters, $C_2-C_5$, achieved comparable performance in contrast with $C_1$ (master cluster). Moreover, cluster $C_3$ achieved a higher performance than $C_1$. This performance enhancement is due to the distillation of knowledge from master to slave clusters during training. The details experiment on the impact of using knowledge distillation is elaborated in Section~\ref{mstt}. Apart from Table~\ref{result2}(a), Table~\ref{result2}(b) illustrates the performance of different clusters on considered datasets after compaction. The results showed a clear margin of improvement in the global and cluster-wise performance while using the clusters compaction in the Fed-RAC. It is due to the increment in the number of participants in each cluster.

\noindent \textit{Observation: An interesting observation from this experiment is that the performance of models in each cluster improves with cluster compaction. Moreover, the improvement in performance is more significant for clusters with a larger number of participants compared to those with fewer participants. }

\begin{table}[h]
\centering
\caption{\textcolor{black}{Impact of cluster compaction on accuracy of the Fed-RAC using different datasets (MNIST, HAR, CIFAR-10, and SHL)}.}
\resizebox{0.48\textwidth}{!}{
\begin{tabular}{|@{}c@{}|c|c|c|c|}
\multicolumn{5}{c}{(a) Accuracy without compaction (cluster count $=5$).} \\ \hline
\multirow{2}{*}{\textbf{Clusters}} & \multicolumn{4}{c|}{\textbf{Accuracy (in \%) on datasets}}                                                                      \\ \cline{2-5} 
                                  & \multicolumn{1}{c|}{\textbf{MNIST}} & \multicolumn{1}{c|}{\textbf{HAR}} & \multicolumn{1}{c|}{\textbf{CIFAR-10}} & \textbf{SHL} \\ \hline
$C_1$   &  $98.58 \pm 0.50$            &  $93.31 \pm0.50$            &  $92.23\pm 0.30$ &  $91.43\pm 0.25$            \\ \hline
$C_2$   &  $98.18 \pm 0.60$            &  $93.68 \pm0.40$            &  $91.55\pm 0.25$ &  $90.17 \pm 0.30$            \\ \hline
$C_3$   &  $\mathbf{98.14 \pm 0.30}$   &  $\mathbf{94.19 \pm 0.20}$  &  $91.23\pm 0.40$ &  $91.76 \pm 0.30$            \\ \hline
$C_4$   &  $97.04 \pm 0.50$            &  $93.55 \pm0.50$            &  $90.58\pm 0.25$ &  $89.88 \pm 0.40$            \\ \hline
$C_5$   &  $96.71 \pm 0.60$            &  $93.01 \pm0.60$            &  $89.45 \pm 0.60$&  $87.82 \pm 0.40$           \\ \hline
\begin{tabular}[c]{@{}c@{}}Global \end{tabular} & $\mathbf{97.73 \pm 0.40}$   & $\mathbf{93.54\pm0.30}$  & $\mathbf{91.01\pm0.45}$  & $\mathbf{90.27\pm0.30}$ \\ \hline
\end{tabular}
}\\

\resizebox{0.48\textwidth}{!}{
\begin{tabular}{|@{}c@{}|c|c|c|c|}
\multicolumn{5}{c}{(b) Accuracy with compaction (clusters count $=4$).} \\ \hline
\multirow{2}{*}{\textbf{Clusters}} & \multicolumn{4}{c|}{\textbf{Accuracy (in \%) on datasets}}                                                                      \\ \cline{2-5} 
                                  & \multicolumn{1}{c|}{\textbf{MNIST}} & \multicolumn{1}{c|}{\textbf{HAR}} & \multicolumn{1}{c|}{\textbf{CIFAR-10}} & \textbf{SHL} \\ \hline
$C_1$   & $98.76 \pm 0.30$ & $93.47 \pm 0.60$  & $92.41\pm 0.40$ & $91.52 \pm 0.30$             \\ \hline
$C_2$   & $98.73 \pm 0.50$ & $93.76 \pm 0.50$  & $92.37\pm 0.25$ & $92.53 \pm 0.40$             \\ \hline
$C_3$   & $98.78 \pm 0.20$ & $94.92 \pm 0.40$  & $92.69\pm 0.35$ & $92.02 \pm 0.40$             \\ \hline
$C_4$   & $98.63 \pm 0.60$ & $94.17 \pm 0.50$  & $91.73\pm 0.30$ & $91.26 \pm 0.35$             \\ \hline
\begin{tabular}[c]{@{}c@{}}Global  \end{tabular} & $\mathbf{98.72 \pm 0.25}$   & $\mathbf{94.08\pm0.35}$  & $\mathbf{92.30\pm0.30}$   & $\mathbf{91.83\pm0.30}$  \\ \hline
\end{tabular}
}\\

\resizebox{0.48\textwidth}{!}{
\begin{tabular}{|@{}c@{}|c|c|c|c|}
\multicolumn{5}{c}{(c) Accuracy with compaction (cluster count $=3$).} \\ \hline
\multirow{2}{*}{\textbf{Clusters}} & \multicolumn{4}{c|}{\textbf{Accuracy (in \%) on datasets}}                                                                      \\ \cline{2-5} 
                                  & \multicolumn{1}{c|}{\textbf{MNIST}} & \multicolumn{1}{c|}{\textbf{HAR}} & \multicolumn{1}{c|}{\textbf{CIFAR-10}} & \textbf{SHL} \\ \hline
$C_1$   & $98.37 \pm 0.45$ & $93.11 \pm 0.40$  & $91.81\pm 0.30$ & $91.22 \pm 0.40$             \\ \hline
$C_2$   & $97.49 \pm 0.35$ & $92.26 \pm 0.25$  & $90.36\pm 0.35$ & $89.77 \pm 0.40$             \\ \hline
$C_3$   & $95.42 \pm 0.40$ & $89.47 \pm 0.35$  & $87.41\pm 0.50$ & $86.82 \pm 0.30$             \\ \hline
\begin{tabular}[c]{@{}c@{}}Global  \end{tabular} & $\mathbf{97.09 \pm 0.40}$   & $\mathbf{91.62\pm0.30}$  & $\mathbf{89.90\pm0.40}$   & $\mathbf{89.27\pm0.35}$  \\ \hline
\end{tabular}
}
\label{result2}

\end{table}

\subsubsection{Impact of communication rounds}
This experiment investigates the impact of different datasets on the convergence of the Fed-RAC and considered baselines. All $40$ participants were involved in the FL operation, and thus FedAvg and FedProx utilized the smallest slave model to ensure deployment and training on all participants. The communication rounds for Fed-RAC were determined as the rounds required for the convergence of the master model plus the maximum rounds required for the convergence of the slowest slave model.

Fig.~\ref{comm} shows the impact of the considered datasets, namely MNIST, HAR, CIFAR-10, and SHL, on the convergence of Fed-RAC, FedAvg, HeteroFL, FedProx, and Oort. The learning curve depicted in the figure displays a classic shape with a two-step behavior. Initially, the performance improves steeply until it reaches a plateau value after some communication rounds. Then, the accuracy increases with more communication rounds. Fed-RAC outperforms the existing approaches on all communication rounds during the experiment. The participants in the master cluster ($C_1$) quickly converge due to sufficient resources to train a large size model. The Fed-RAC approach also incorporates KD to train the models at the participants, leading to well-behaved optimization steps compared to non-KD based training and reduced communication rounds. On the MNIST dataset, all approaches achieved convergence at lower rounds with marginal improvement afterwards, as shown in Fig.~\ref{comm}. This is due to the balanced and sufficient number of instances for all classes in MNIST. FedAvg achieved slower convergence with minimal accuracy due to incompetence in handling heterogeneity among the participants and using a small size model to accommodate all 40 participants during training. HeteroFL achieved comparable performance to the Fed-RAC due to the strategy in addressing heterogeneity. 

\noindent \textit{Observation: Firstly, the convergence curve of training exhibits a typical trend, with a steep increase in performance at the start, followed by a gradual plateau. Secondly, KD  has a significant impact on the model's performance. }

\begin{figure}[h]
\centering
\begin{tabular}{cc}
\hspace{-0.2cm} \includegraphics[scale=0.40]{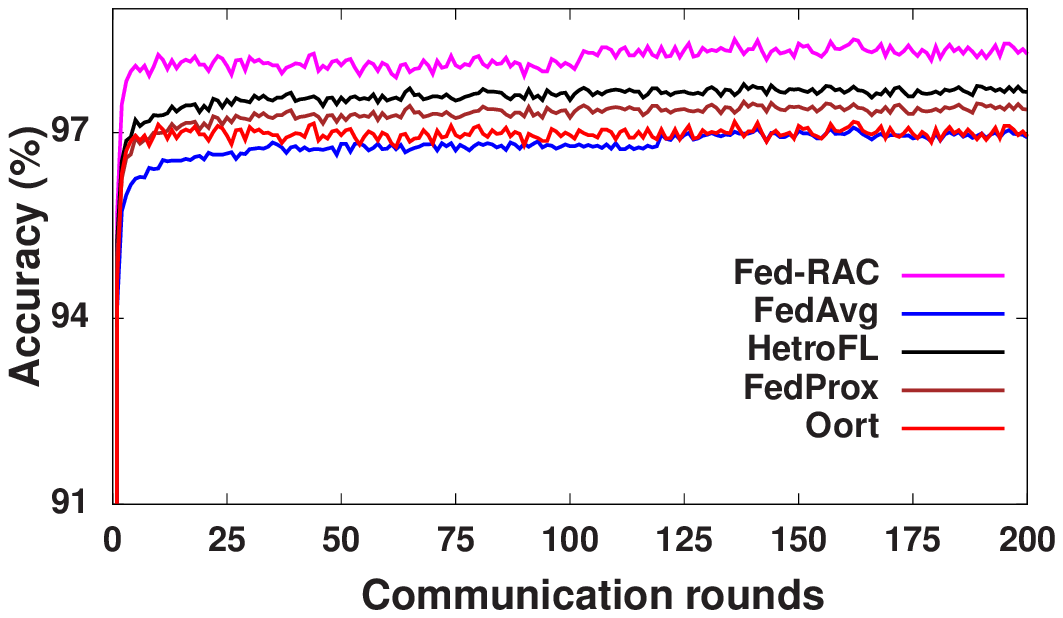} & \hspace{-0.25cm}\includegraphics[scale=0.40]{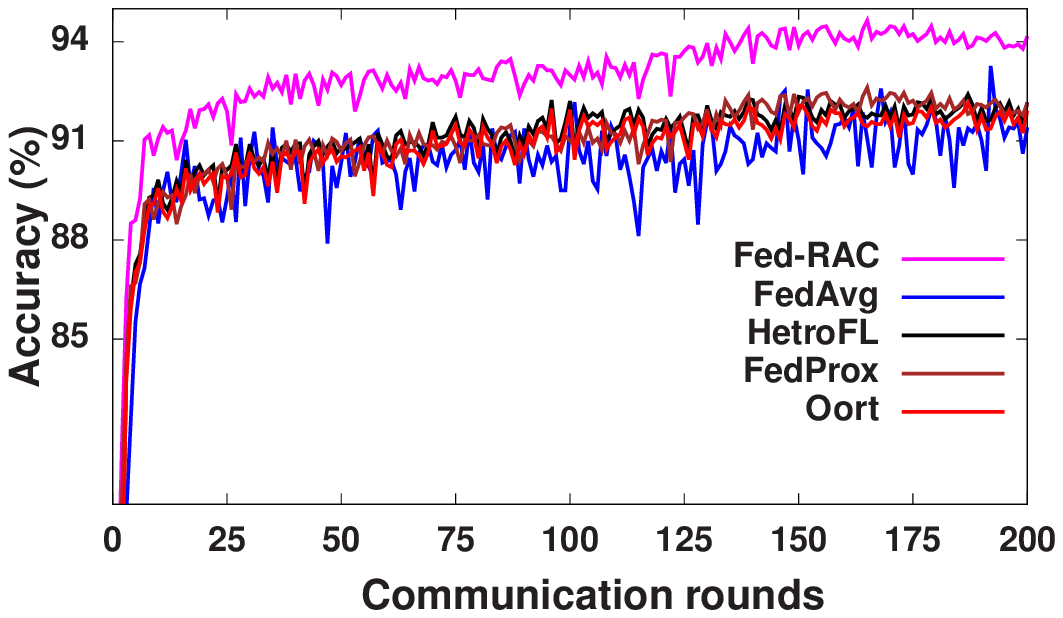} \\
\scriptsize{(a) MNIST dataset.} & \scriptsize{(b) HAR dataset.}\\ 
\hspace{-0.2cm} \includegraphics[scale=0.40]{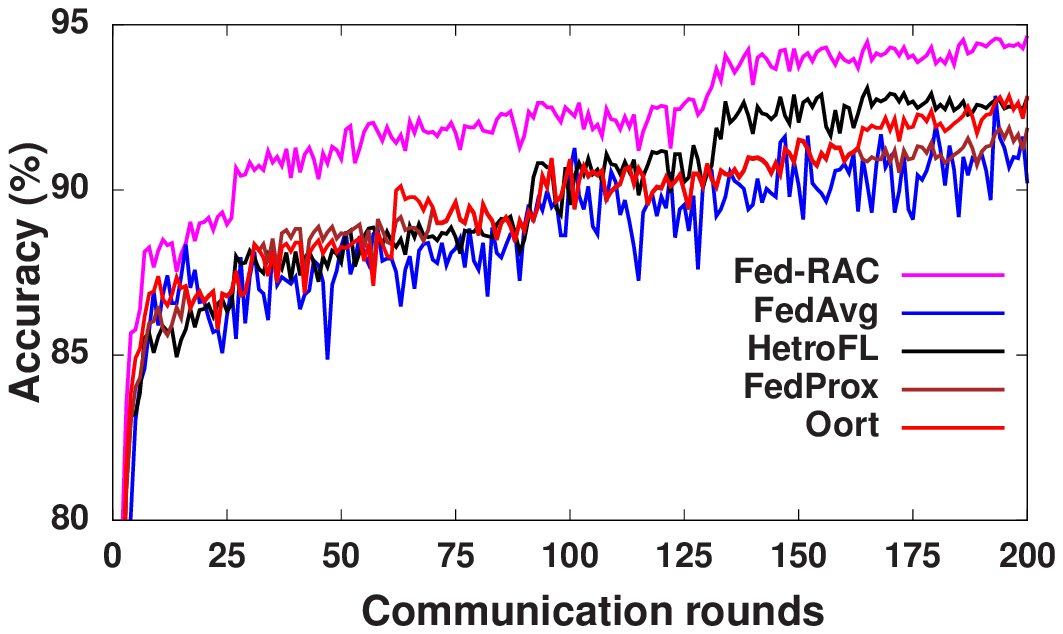} & \hspace{-0.25cm}\includegraphics[scale=0.40]{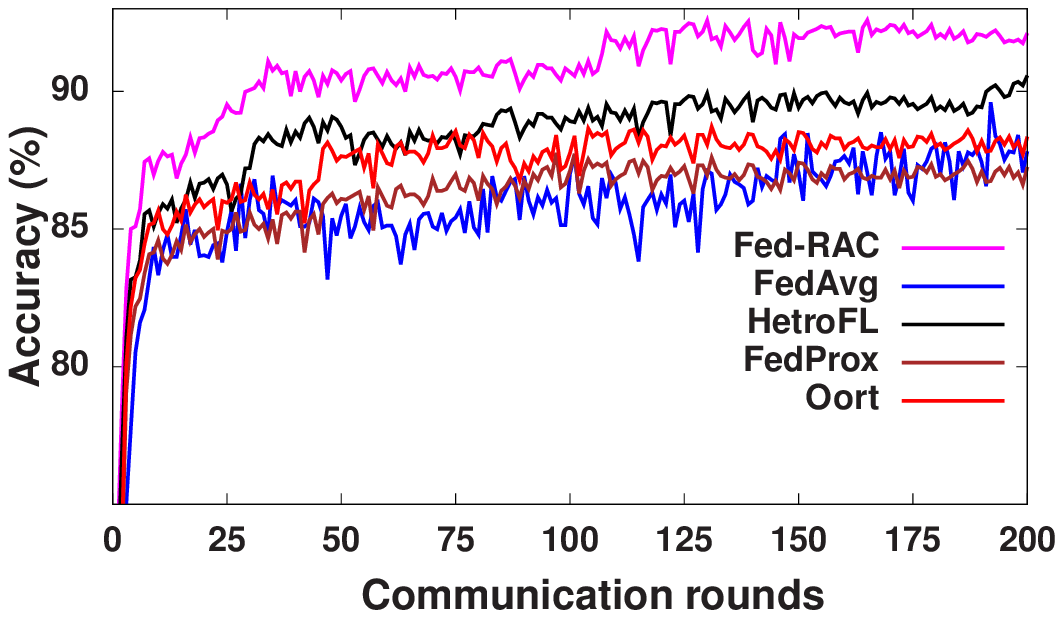} \\
\scriptsize{(c) CIFAR-10 dataset.} & \scriptsize{(d) SHL dataset.}
\end{tabular}

         \caption{\textcolor{black}{Illustration of impact of datasets on the convergence rounds of Fed-RAC, FedAvg, HeteroFL, FedProx, and Oort.}} 
         \label{comm}
 \end{figure}

\subsubsection{Impact of master-slave technique}\label{mstt}
In this experiment, we aim to evaluate the performance improvement of the slave models assigned to each cluster (other than the master cluster) using the master-slave technique discussed in Section~\ref{mspi}. We consider the four clusters, $C_1-C_4$, obtained from the compaction in the previous result. The communication round is fixed at $200$. However, to ensure brevity, we only present the results on HAR and CIFAR-10.

Fig.~\ref{rmst} illustrates the impact of the master slave technique on the performance of models in different slave clusters. Clusters $C_2-C_4$ gain significant improvement in performance due to the distillation of knowledge from the master model in $C_1$, as shown in Fig.~\ref{rmst}(b) and Fig.~\ref{rmst}(d). The results demonstrated that the improvement in the model's performance is significant at low resource clusters ($C_4$) and reduced gradually to $C_2$. It is because if the size of the cluster model is small then the logits difference between master and slave is higher. Contrarily, if the difference between the size of the cluster model and the master model is less, the logit difference is limited; thus, the performance gain is low. Cluster $C_4$ gains accuracy of $\approx 8\%$ for HAR and $\approx 11\%$ for CIFAR-10 datasets, whereas the performance  gain for $C_2$ is $\approx 2\%$ for both datasets. Furthermore, in FL-based training, we considered participants with heterogenous resources; thus, participants with the highest and lowest resources, respectively, achieved colossal and most minor performances. It also creates a significant difference between the performance of the models in the largest and smallest clusters, which aggregately results in performance compromise despite clustering. Therefore, KD is incorporated to enhance the performance of models in the smaller clusters. 
 
\noindent \textit{Observation: An interesting observation from our experiment is that the master-slave technique leads to a significant improvement in the performance in the slave clusters. Furthermore, the extent of improvement in the performance of the slave clusters depends on their respective model sizes.}

\begin{figure}[h]
\centering
\begin{tabular}{cc}
\hspace{-0.2cm} \includegraphics[scale=0.40]{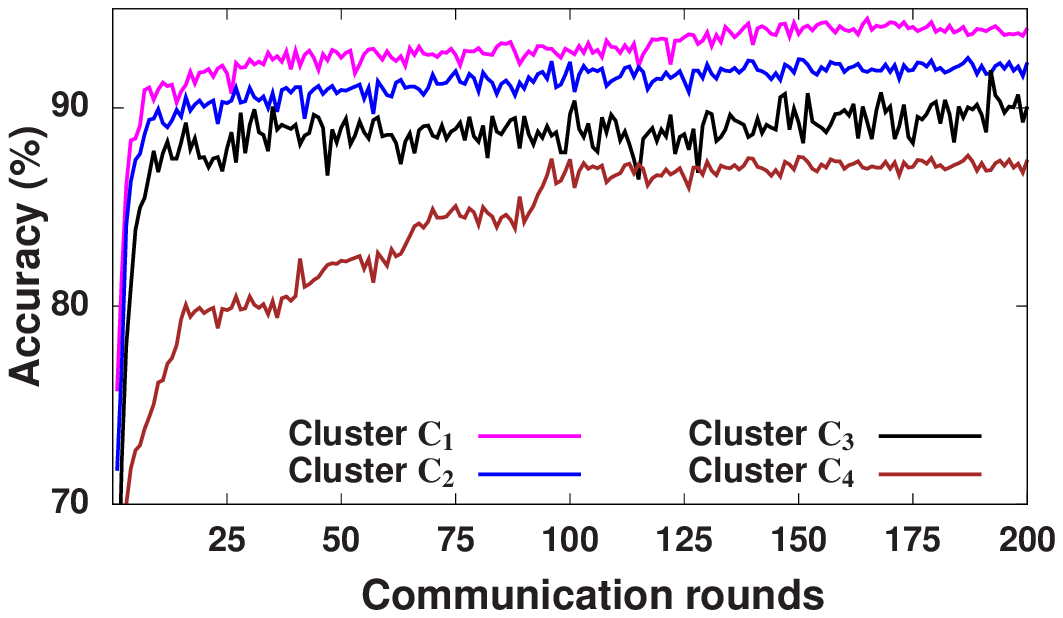} & \hspace{-0.25cm}\includegraphics[scale=0.40]{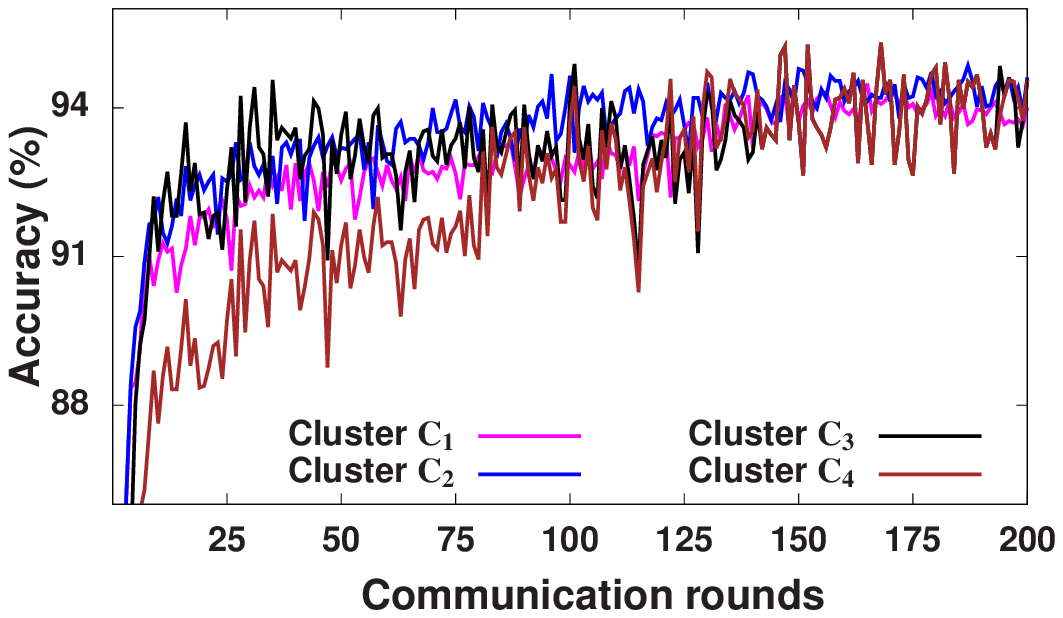} \\
\scriptsize{(a) HAR without KD.} & \scriptsize{(b) HAR with KD.}\\
\hspace{-0.2cm} \includegraphics[scale=0.40]{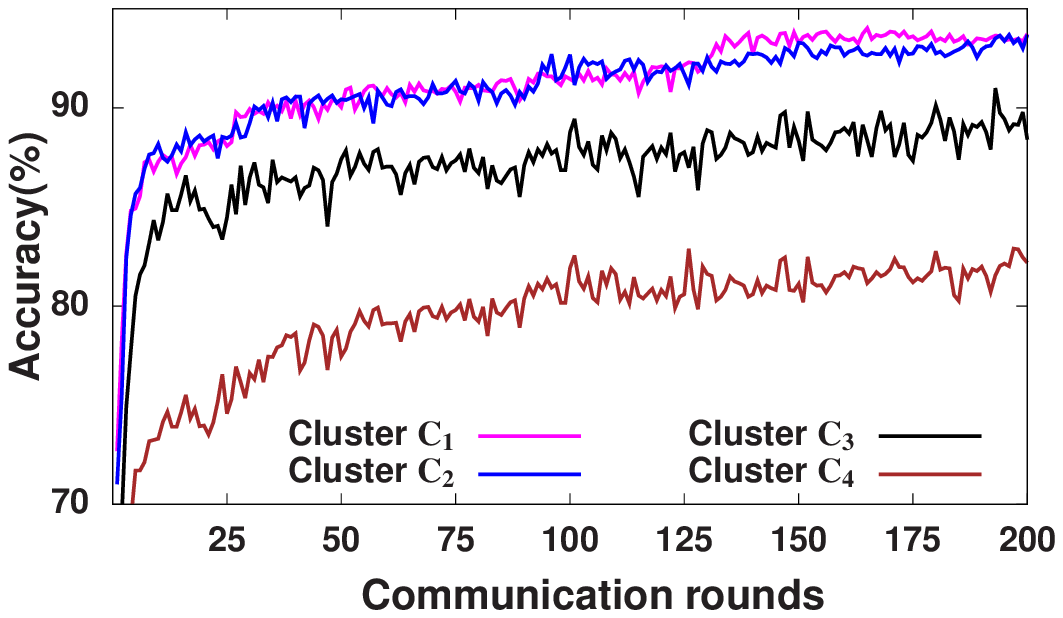} & \hspace{-0.25cm}\includegraphics[scale=0.40]{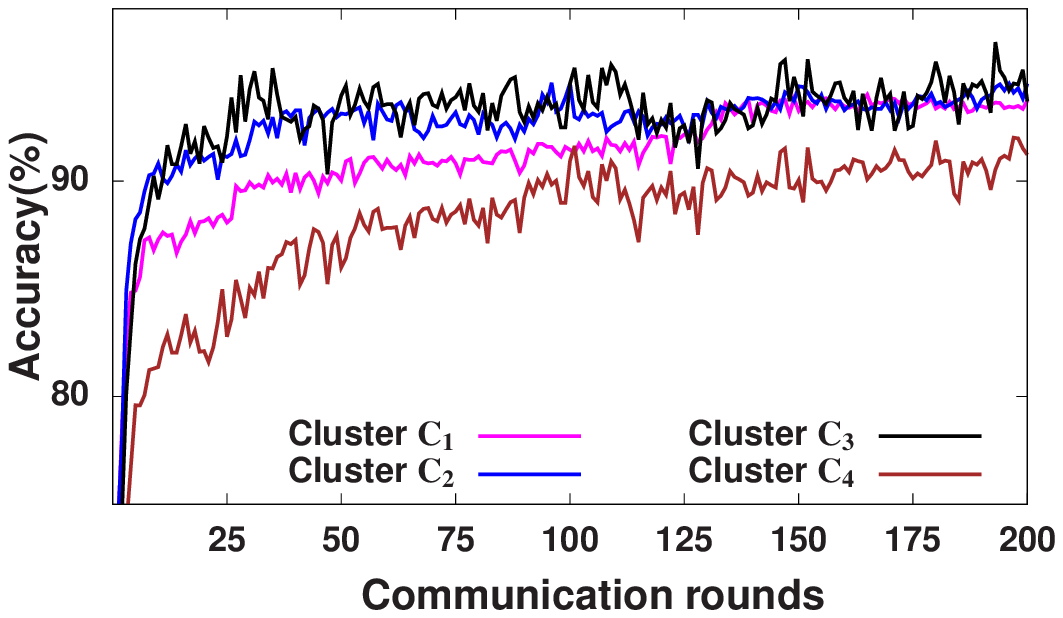} \\
\scriptsize{(c) CIFAR-10 without KD.} & \scriptsize{(d) CIFAR-10 with KD.}
\end{tabular}

         \caption{Impact of master slave technique on the performance of models in different clusters using HAR and CIFAR-10 datasets.} 
         \label{rmst}
 \end{figure}

\subsubsection{Impact of rounds-to-reach x\%}
The objective of this experiment is to investigate the effectiveness of the proposed Fed-RAC in achieving a global accuracy of $x\%$ within a certain number of communication rounds. To achieve this, we have set the value of $x$ to be $96$, $92$, $88$, and $85$ for MNIST, HAR, CIFAR-10, and SHL datasets, respectively, taking into account the convergence rates of these datasets. Fed-RAC involves training the model in the master cluster followed by parallel training of models in the slave clusters. As such, we define the Total Required Rounds (TRR) for complete training as the sum of rounds required to train the model in the master cluster ($C_1$) and the maximum rounds required to train the model in any of the slave clusters ($\max rounds \{C_2, C_3, C_4\}$).

Table~\ref{xp} presents the results of the rounds-to-reach $x\%$ performance metric on the considered datasets and illustrates the impact of this metric on the Fed-RAC approach. The results indicate that the Fed-RAC approach (cluster-wise with KD) outperforms the baseline approaches, including cluster-wise without KD. This can be attributed to two main reasons. Firstly, the participants in the master cluster ($C_1$) have sufficient resources to train large models, which leads to quicker convergence. Secondly, the Fed-RAC approach incorporates KD to train the models at the participants, resulting in well-behaved optimization steps compared to non-KD.

Regarding the convergence of \textit{cluster-wise without KD}, the results are not reported for models in clusters $C_3$ and $C_4$ on HAR, CIFAR-10, and SHL datasets. This is because, in the absence of KD, the participants in clusters $C_3$ and $C_4$ are unable to achieve the desired $x\%$ accuracy within the cap of $200$ communication rounds. Furthermore, we used small models in FedAvg, Oort, and FedProx to involve all $40$ participants. Although the use of KD appears to incur higher computational costs compared to the baselines that do not incorporate KD, Fed-RAC achieves the desired performance in fewer communication rounds, thus reducing the overall computational cost. Moreover, the number of local epochs required for convergence decreases with the cluster size, which also reduces the computational cost in Fed-RAC.

\noindent \textit{Observation: We observe that KD from the large-size model to the lightweight model not only improves the performance but also reduces the communication rounds for convergence.} 

\begin{table}[h]
\centering
\caption{Illustration of impact of rounds-to-reach $x\%$ global accuracy on considered datasets. TRR=Total Required Rounds= rounds($C_1$)+$\max$ rounds \{$C_2$, $C_3$, $C_4$\}, A1= FedAvg, A2=HetroFL, A3=FedProx, and A4=Oort}.

\resizebox{0.50\textwidth}{!}{
\begin{tabular}{|@{}c@{}|c|c@{}|c@{}|c@{}|c@{}|@{}c@{}|c@{}|c@{}|c@{}|c@{}|@{}c@{}|c@{}|c@{}|c@{}|c@{}|}
\hline
\multirow{3}{*}{\textbf{Dataset}} & \multirow{3}{*}{$\mathbf{x\%}$} & \multicolumn{10}{c|}{\textbf{Fed-RAC (proposed)}}                                                                                                                                                                                                  & \textbf{A1} & \textbf{A2} & \textbf{A3} & \textbf{A4} \\ \cline{3-12}
&                               & \multicolumn{5}{c|}{\textbf{Cluster w/o KD}}                                                                                     & \multicolumn{5}{c|}{\textbf{Cluster w KD}}  &      &      &      &                                \\ \cline{3-12}

&      & $C_1$ & $C_2$ & $C_3$ & $C_4$ & TRR & $C_1$ & $C_2$ & $C_3$ & $C_4$ & TRR &     &    &    &                             \\ \hline

\textbf{MNIST}      & $96$   & $2$  & $2$  & $5$  & $9$  & $11$  & $2$  & $2$  & $3$  & $5$   & $\mathbf{7}$      & $9$   & $11$     &   $8$      & $9$                                    \\ \hline

\textbf{HAR}        & $92$   & $36$   & $47$   &  -  &  -  &  $83$    & $36$  &  $17$  & $29$   & $41$  &    $\mathbf{77}$   &  $92$    & $102$    & $89$      &  $86$        \\ \hline

\textbf{CIFAR-10}   & $88$   & $51$   & $59$   & -  & -  & $110$     &  $51$ & $23$  & $37$   & $53$  &  $\mathbf{104}$     &  $112$    & $121$ & $108$                                                       &  $115$                                                   \\ \hline

\textbf{SHL}        & $86$   &  $67$  & $74$  &  -  &  -  &   $141$   &  $67$  &  $34$ & $39$   & $61$   &  $\mathbf{128}$      &  $137$                                   &   $146$ &  $135$  & $137$                                                    \\ \hline
\end{tabular}
}
\label{xp}
\end{table}

\subsubsection{Leave-one-out}
The objective of this experiment was to assess the overall performance of Fed-RAC and several baseline approaches in a scenario where instances of a randomly selected class label were not included in the training but appeared in the testing. The class label with the highest number of instances was selected as the leave-out class during the experiment. The communication rounds were set to $200$, and the parameters and local epochs were determined according to the implementation details discussed in Section~\ref{implementation}.

In Fig.~\ref{leave}, the impact of removing instances of one class label from the training of all participants in FL is demonstrated. The results show that Fed-RAC outperforms the existing approaches, which is consistent with the performance pattern observed in previous results. The approach that does not use KD clustering (referred to as the "without KD clustering approach") achieved the lowest performance, likely due to the small size models trained on slave clusters with a limited number of participants in each cluster. This negatively impacted the overall performance of the approach. The MNIST dataset achieved the highest performance due to the large number of instances for classes other than the excluded one. Conversely, the SHL dataset had the lowest performance due to the excluded class having the highest number of instances.

\noindent \textit{Observation: An interesting observation is that excluding training instances for certain class labels leads to a deterioration in performance. The decline in performance is more pronounced when a class with a large number of instances is excluded. Furthermore, the absence of KD results in a more rapid degradation in performance compared to the KD.} 

\begin{figure}[h]
\centering
\includegraphics[scale=0.56]{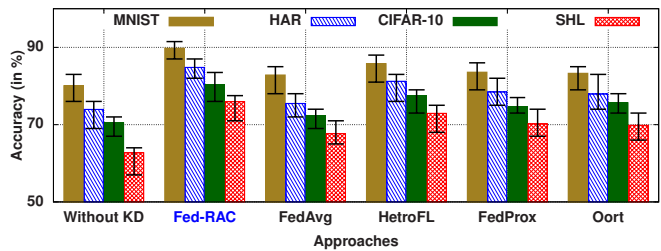}
\caption{Impact of leave-one-out test metric on MNIST, HAR, CIFAR-10, and SHL datasets using Fed-RAC (without KD), Fed-RAC (with KD), FedAvg, HeteroFL, FedProx, and Oort approaches.}
\label{leave}
\end{figure}

\subsubsection{Learning rate}
This experiment aimed to investigate how the learning rate affects the performance of Fed-RAC. MNIST, HAR, CIFAR-10, and SHL datasets were used, and the communication rounds were set to $5$, $10$, $20$, and $20$, respectively. The rounds were restricted as the approach converged at any learning rate at higher communication rounds.

Table~\ref{lr} depicts the impact of distinct learning rates on the accuracy of model in the master cluster of the Fed-RAC using MNIST, HAR, CIFAR-10, and SHL datasets. The results demonstrated the efficacy of the Fed-RAC on a smaller learning rate (\textit{e.g.,} $0.002$). We obtained the lowest accuracy for the learning rate of $0.010$ due to faster convergence. The model converged sub-optimally at higher learning rate; thus, it suffered from performance compromise. Fed-RAC converged faster for the MNIST; hence, we achieved accuracy beyond 90\% for all the datasets at different learning rates only at $5$ communication rounds. The achieved accuracy of the Fed-RAC approach follows a linear pattern for all the datasets; however, we also observed plateaued behavior for learning rates between $0.006$ to $0.008$. Additionally, the difference between cluster accuracy at the learning rate of $0.002$ and $0.010$ is more than $8\%$, which signifies the importance of selecting an optimal learning rate during training.\\ 
\noindent \textit{Observation: The results indicated that the learning rate is a critical factor in achieving higher performance. A smaller learning rate is beneficial for longer communication rounds, while a larger learning rate is better for shorter rounds.} 

\begin{table}[h]
\centering
\caption{Impact of learning rate on the accuracy of the model in master cluster. CR=Communication rounds.}

\resizebox{0.49\textwidth}{!}{
\begin{tabular}{|c|c|c|c|c|c|c|}
\hline
\multicolumn{1}{|c|}{\multirow{2}{*}{\textbf{Datasets}}} & \multicolumn{1}{c|}{\multirow{2}{*}{\textbf{\begin{tabular}[c]{@{}c@{}}CR\end{tabular}}}} & \multicolumn{5}{c|}{\textbf{Accuracy (in \%) on learning rates}}                                                                                                                                                                                                                                                                                                                                              \\ \cline{3-7} 
        &                  & $\mathbf{0.002}$  & $\mathbf{0.004}$     & $\mathbf{0.006}$  & $\mathbf{0.008}$  & $\mathbf{0.010}$ \\ \hline
\textbf{MNIST}    &  $5$   & $98.07$           & $96.44$  & $93.32$  & $92.97$                      & $90.37$ \\ \hline
\textbf{HAR}      &  $10$  & $89.93$           & $87.24$  & $86.05$  & $83.75$                       & $79.24$  \\ \hline
\textbf{CIFAR-10} &  $20$  & $84.41$           & $83.73$  & $81.29$  & $80.73$                      & $77.12$ \\ \hline
\textbf{SHL}      &  $20$  & $82.14$           & $80.71$  & $79.64$  & $79.32$                    & $74.23$ \\ \hline
\end{tabular}
}
\label{lr}
\end{table}

\section{Discussion and future work}\label{discuss}
In this section, various issues are discussed that need to be addressed in future work in conjunction with the proposed approach. The approach uses a master-slave technique where logits from the master cluster model are sent to the remaining clusters. However, this could potentially expose private training data or enable participants to reconstruct models. To address these privacy concerns, future work on incorporating security aspects is necessary, as motivated by previous work on differential privacy in FL~\cite{differentail}. Furthermore, while Fed-RAC considers participant heterogeneity, it does not account for noise in data instances and labels. Therefore, future work will involve incorporating such noise in the model training process. Additionally, the approach independently trains a local model for each cluster without leveraging information from models in other clusters, except for logit vectors from the master cluster model. To address this limitation, future work will focus on developing mechanisms for aggregating information on trained models from different clusters.

\section{Conclusion}\label{conc}
In this paper, a federated learning approach called Fed-RAC is proposed to address the negative impact of heterogeneous participants. Unlike previous studies, Fed-RAC trains local models on all participants despite differences in heterogeneity and training time. The approach first identifies the optimal number of clusters based on available devices and networking resources, then generates and randomly initializes a model that is used for compression to obtain models for all clusters. A participant assignment mechanism and a master-slave technique are introduced to improve the performance of lightweight models using knowledge distillation. Experimental evaluation is conducted to verify the approach's effectiveness on existing datasets, leading to several key findings: successful federated learning requires proper management of participant heterogeneity, resource-aware clustering helps identify the optimal number of clusters, the number of data instances significantly affects cluster performance, and the master-slave technique enhances performance based on model size.

\bibliographystyle{IEEEtran}
\bibliography{lpencil}

\end{document}